\documentclass[11pt,english]{article}
\usepackage{lmodern}
\usepackage{hyperref}

\usepackage[T1]{fontenc}
\usepackage[utf8]{inputenc}
\usepackage{geometry}
\usepackage{color}
\usepackage{float}
\usepackage{amsmath}
\usepackage{amssymb}
\usepackage{graphicx}
\usepackage{setspace}

\hypersetup{colorlinks=true,linkcolor=black}

\makeatletter

\providecommand{\tabularnewline}{\\}

\usepackage[winfonts,UTF8]{ctex}
\usepackage{slashed}
\usepackage{cite}
\date{}

\makeatother

\usepackage{babel}
\begin{document}

\title{Holographic subregion complexity under a thermal quench}

\author{Bin Chen$^{1,2,3}$, Wen-Ming Li$^{2}$, Run-Qiu Yang$^{4}$, Cheng-Yong
Zhang$^{1}$, Shao-Jun Zhang$^{5}$\thanks{bchen01@pku.edu.cn, liwmpku@pku.edu.cn, aqiu@kias.re.kr, zhangcy0710@pku.edu.cn,
sjzhang84@hotmail.com}}
\maketitle
\begin{center}
\textsl{$^{1}$Center for High Energy Physics, Peking University,
No.5 Yiheyuan Rd, Beijing 100871, P. R. China}\\
\textsl{$^{2}$Department of Physics and State Key Laboratory of Nuclear
Physics and Technology, Peking University, No.5 Yiheyuan Rd, Beijing
100871, P.R. China}\\
\textsl{$^{3}$Collaborative Innovation Center of Quantum Matter,
No.5 Yiheyuan Rd, Beijing 100871, P. R. China}\\
\textsl{$^{4}$Quantum Universe Center, Korea Institute for Advanced
Study, Seoul 130-722, Korea}\\
\textsl{$^{5}$Institute for Advanced Physics and Mathematics, Zhejiang
University of Technology, Hangzhou 310023, China}
\par\end{center}
\begin{abstract}
We study the evolution of holographic subregion complexity under
a thermal quench in this paper. From the subregion CV proposal in the AdS/CFT correspondence, the  subregion complexity in the CFT is holographically
captured by the volume of the codimension-one surface enclosed by the codimension-two extremal entanglement
surface and the boundary subregion. Under a thermal quench, the dual gravitational configuration is
described by a Vaidya-AdS spacetime. In this case we find that the holographic subregion
complexity always increases at early time, and after reaching a maximum it   decreases and gets to saturation.
Moreover we notice that when the size of the strip is large enough and the quench is fast enough, in $AdS_{d+1}(d\geq3)$ spacetime the evolution of the complexity is discontinuous and there is
 a sudden drop due to the transition of the extremal entanglement surface.
We discuss the effects of the quench speed, the strip size, the black hole mass and the spacetime dimension on the evolution of the subregion complexity  in detail numerically.
\end{abstract}
\newpage
\tableofcontents{}

\section{Introduction}

The interplay between the black hole physics and quantum information has a long history.
One of the most interesting concern is on the physics of the black hole horizons.
Few years ago, L. Susskind pointed out that just considering the entanglement was not enough to
understand  the horizon\cite{Susskind2014,Susskind2014ee}. In particular he  argued that the creation of the firewall behind the horizon
was actually a problem of computational complexity in the framework of ER$=$EPR\cite{Maldacena:2013xja}, and furthermore proposed that the
complexity could be read by  the volume of a worm hole in the bulk. Since then, the complexity in quantum field theory and gravity has been discussed intensely.

Complexity  is an important conception in the information theory. In the quantum circuit
model, it measures how many minimum simple gates are
needed to complete a given task transferring a reference state to a
target state \cite{Watrous2008,Osborne2011,Gharibian2014}. However,
this manipulation can not directly generalized to quantum field theory
due to the ambiguity in defining the simple operation and the
reference state in a system of infinite degrees of freedom. There have been some attempts trying to give
a well-defined  complexity in quantum field theory\cite{Nielsen2005,Nielsen2006,Nielsen2007,Mayers2017,Chapman1707,Yang2017,Khan2018,Caputa2017,Caputa2017b,Yang1803}.

From the AdS/CFT correspondence, there have been two different proposals on holographic complexity, which are referred to
as the CV (Complexity=Volume) conjecture \cite{Susskind2014,Susskind2014ee,Stanford2014,Susskind2014sb}
and the CA (Complexity=Action) conjecture \cite{Brown2015,Brown2015b}
respectively. The CV conjecture states that the complexity of a boundary
state on a time slice $\Sigma$ is dual to the extremal volume of
the corresponding codimension-one hypersurface $\mathcal{B}$ whose
boundary is anchored at $\Sigma$:
\begin{equation}
C_{V}=\underset{\Sigma=\partial\mathcal{B}}{\mathbf{max}}\left(\frac{V(\mathcal{B})}{G_{d+1}l}\right).
\end{equation}
Here $G_{d+1}$ is the gravitational constant in $AdS_{d+1}$ and
$l$ is some length scale associated with the bulk geometry, e.g.
the anti-de Sitter (AdS) curvature scale or the radius of a black
hole. The ambiguity in the length scale $l$ is unsatisfactory so that the CA conjecture was proposed.  In the
CA conjecture  the complexity of the boundary state is identified holographically with the
gravitational action evaluated on the Wheeler-DeWitt (WDW) patch in
the bulk:
\begin{equation}
C_{A}=\frac{I_{WDW}}{\pi\hbar}.
\end{equation}
The WDW patch is the bulk domain of dependence of a bulk Cauchy slice
anchored at the boundary. It is the causal domain of the hypersurface
$\Sigma$ defined in the CV conjecture. Both CV and CA satisfy important
requirements on the complexity such as the Lloyd's bound \cite{Lloyd2000,Yang2016,Cai2016,Cai2017,An2018}.

Like the entanglement entropy, it is also interesting to consider
the complexity of a subregion. Instead of a pure state in the whole
boundary,  it is generally a mixed state
produced by reducing the boundary state to a specific subregion (donated
by $\mathcal{A}$). Since the mixed state
is encoded in the entanglement wedge in the bulk \cite{Czech1204,Headrick1408},
the subregion complexity should involve the entanglement wedge. In \cite{Mayers1612}
and \cite{Alishahiha2015}  the CA and CV proposals have been generalized to
the subregion situation respectively. For the subregion version of the
CA proposal,  the complexity of subregion $\mathcal{A}$  equals
 the action of the intersection of the WDW patch and the entanglement
wedge \cite{Mayers1612}. As for the subregion CV proposal, the complexity
equals the volume of the extremal hypersurface $\Gamma_{\mathcal{A}}$
enclosed by the boundary subregion $\mathcal{A}$ and corresponding
Ryu-Takayanagi(RT) surface $\gamma_{\mathcal{A}}$ \cite{RT2006,RT2006b,HRT1509,HRT1607}.
Precisely, it can be computed by
\begin{equation}
C_{\mathcal{A}}=\frac{V(\Gamma_{\mathcal{A}})}{8\pi RG_{d+1}}
\end{equation}
 where $R$ is the AdS radius. It has been suggested that the possible
dual field theory quantity is the fidelity susceptibility in quantum
information theory \cite{Miyaji1025,Alishahiha2015}. 

The subregion CV proposal can be understood intuitively from the
entanglement renormalization \cite{Vidal2005,Swingle2009} and the tensor
network \cite{Vidal1412,Vidal1502}. The entanglement entropy can
be estimated by the minimal number of bonds cut along a curve which
is reminiscent of the entanglement curve. Then the holographic complexity
can be estimated by the number of nodes in the area enclosed by the
curve cutting the bonds \cite{Stanford2014}. This idea becomes more
transparent from the surface/state correspondence conjecture \cite{Takyanagi1503}
in which the complexity between two states is proportional to the number
of operators enclosed by the surface corresponding to the target state
and the surface corresponding to the reference state. Obviously the
complexity is proportional to the volume enclosed by these two surfaces.
This picture has been described in \cite{Caputa2017,Caputa2017b}
and also in \cite{Czech1706}. For other works
 on the subregion complexity, please see \cite{subBenAmi2016,subRoy2017,subBanerjee2017,subBakhshaei2017,subSarkar2017,subZangeneh2017,
 subMomeni2017,subRoy2017b,subCarmi2017,Abt2017,Du2018}.

In this paper, we would like to study the subregion complexity in a time-dependent background {using the subregion CV conjecture}.
In particular we compute the evolution of the subregion complexity after a global thermal quench in detail.
The quenched system has been viewed as an effective model to study
thermalization both in field theory and holography \cite{Calabrese2005,AbajoArrastia2010,Albash2010,Balasubramanian1103,Liu1311}.
On the gravity side, such a quench process in a conformal field theory (CFT) is described by the
process of black hole formation due to the gravitational collapse of
a thin shell of null matter, which in turn can be described by a Vaidya
metric.
{
The pure state complexity in the same background has been studied analytically under the
condition that the shell is pretty thin in \cite{Chapman:2018}. It was found that  the growth of the complexity is just the same as that for eternal black hole at the late time.
}



This paper is organized as follows. In Sec.2, we introduce the framework
 to evaluate the subregion complexity. In Sec.3, we study holographically the evolution of the complexity
after a thermal quench in detail. We summarize our results in Sec.4.

\section{General framework}

In this section, we introduce the general framework to study the subregion complexity in the time-dependent background corresponding to a thermal-quenched  CFT. A thermal quench in a CFT can be described holographically
by the collapsing of a thin shell of null dust  falling from the AdS
boundary to form a black hole. This process can be modeled by a Vaidya-AdS
metric. The metric of the Vaidya-$AdS_{d+1}$ spacetime
with a planar horizon can be written in terms of the Poincare coordinate
\begin{align}
&ds^{2}=  \frac{1}{z^{2}}\left[-f(v,z)dv^{2}-2dzdv+dx^{2}+\sum_{i=1}^{d-2}dy_{i}^{2}\right],\label{eq:VaidyaMetric}\\
&f(v,z)=  1-m(v)z^{d}.\nonumber
\end{align}
 In the present work, the $AdS$ space radius is rescaled to be unit  such that all the
coordinates are dimensionless. In (\ref{eq:VaidyaMetric}) the coordinate $v$ labels the ingoing
null trajectory and coincides with the time coordinate $t$ on the
boundary $z\to0$. $m(v)$ is the mass function of the in-falling shell.
In the following, we will take it to be of the form
\begin{equation}
m(v)=\frac{M}{2}\left(1+\tanh\frac{v}{v_{0}}\right)\label{eq:Massfunction}
\end{equation}
 where $v_{0}$ characterizes the thickness of the shell, or the time over which the quench occurs. Actually  the quench could be taken approximately as starting at $-2v_0$ and ending at $2v_0$. With this mass function, the Vaidya metric interpolates between
a pure AdS in the limit $v\to-\infty$ and a Schwarzschild-AdS (SAdS)
black hole with mass $M$ in the limit $v\to\infty$. When $v_{0}$ goes to zero, the spacetime is simply the joint of a pure
AdS and a SAdS at $v_{0}=0$. The  apparent
horizon in the Vaidya-AdS spacetime locates at
\begin{equation}
r_{h}=m(v)^{-1/d}.\label{eq:horizon}
\end{equation}

\begin{figure}
\begin{centering}
\begin{tabular}{c}
\includegraphics[scale=0.7]{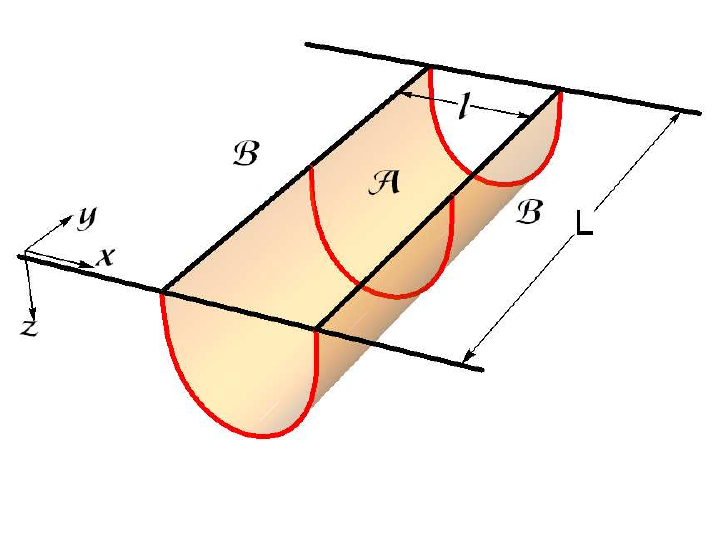}\tabularnewline
\end{tabular}
\par\end{centering}
\caption{\label{fig:StripStatic}The profile of the minimal surface of an infinite
strip in pure $AdS$ background.}
\end{figure}

We consider the subregion of  an infinite strip $\mathcal{A}=x\in\left(-\frac{l}{2},\frac{l}{2}\right),y_i\in\left(-\frac{L}{2},\frac{L}{2}\right)$
with $L\to\infty$ and finite $l$. The profile of the strip in a static AdS background
is shown in Fig. \ref{fig:StripStatic}. We are going to study the evolution of the subregion complexity of the strip holographically in the Vaidya-AdS spacetime.

As proposed in \cite{Alishahiha2015}, the subregion complexity in
a static background is proportional to the volume of a codimension-one
time slice in the bulk geometry enclosed by the boundary region and
the corresponding extremal codimension-two Ryu-Takayanagi (RT) surface. This proposal can be generalized to the dynamical
spacetime. For a subregion $\mathcal{A}$ on the boundary, its holographic
entanglement entropy (HEE) is captured by a codimension-two bulk surface
with vanishing expansion of geodesics \cite{HRT1509}, i.e., the  Hubeny-Rangamani-Takayanagi (HRT) surface $\gamma_{\mathcal{A}}$. The corresponding
subregion complexity is then proportional to the volume of a codimension-one hypersurface
$\Gamma_{\mathcal{A}}$ which takes $\mathcal{A}$
and $\gamma_{\mathcal{A}}$ as boundaries.
Note that $\gamma_{\mathcal{A}}$ and hence $\Gamma_{\mathcal{A}}$
do not live on a constant time slice in general for a dynamical background.
To get the corresponding subregion complexity, we need work out the
corresponding extremal codimension-two surface $\gamma_{\mathcal{A}}$
first.

\subsection{Holographic entanglement entropy }

Due to the symmetry of the strip, the corresponding extremal surface $\gamma_{\mathcal{A}}$
in the bulk can be parametrized as
\begin{equation}
v=v(x),\ \ z=z(x),\ \ z(\pm l/2)=\epsilon,\ \ v(\pm l/2)=t-\epsilon,
\end{equation}
 where $\epsilon$ is a cut-off. The induced metric on the extremal
surface is
\begin{equation}
ds^{2}=\frac{1}{z^{2}}\left[-f(v,z)v'^{2}-2z'v'+1\right]dx^{2}+\frac{1}{z^{2}}\sum_{i=1}^{d-2}dy_{i}^{2}.
\end{equation}
 The area is
\begin{equation}
Area(\gamma_{A})=L^{d-2}\int_{-l/2}^{l/2}\frac{\sqrt{1-f(v,z)v'^{2}-2z'v'}}{z^{d-1}}dx,
\end{equation}
 where $L$ is the length along the spatial directions $y_{i}$. Since the Lagrangian
\begin{equation}
\mathcal{L}_{S}=\frac{\sqrt{1-f(v,z)v'^{2}-2z'v'}}{z^{d-1}}
\end{equation}
 does not depend on $x$ explicitly, the Hamiltonian is conserved.
\begin{equation}
\mathcal{H}_{S}=\frac{1}{z^{d-1}\sqrt{1-f(v,z)v'^{2}-2z'v'}}.
\end{equation}
 Due to the symmetry of the strip, there is a turning point of the extremal
surface $\gamma_{\mathcal{A}}$ locating at $x=0$. At this point
we have
\begin{equation}
v'(0)=z'(0)=0,\ \ z(0)=z_{\ast},\ \ v(0)=v_{\ast},
\end{equation}
 where $z_{\ast},v_{\ast}$ are two parameters that characterize the
extremal surface. The constant Hamiltonian then gives
\begin{equation}
1-f(v,z)v'^{2}-2z'v'=\frac{z_{\ast}^{2d-2}}{z^{2d-2}}.\label{eq:RTt}
\end{equation}
Taking the derivative (\ref{eq:RTt}) with respect to $x$ and using the equation
of motion for $z(x)$, we get
\begin{equation}
-2(d-1)+2zv''+v'\left[2(d-1)f(v,x)v'+4(d-1)z'-zv'\partial_{z}f(v,z)\right]=0.\label{eq:vpp}
\end{equation}
 Taking the derivative (\ref{eq:RTt}) with respect to $x$ and using the equation
of motion for $v(x)$, we get
\begin{align}
0= & 2(d-1)f(v,z)^{2}v'^{2}+f(v,z)\left[-2(d-1)+4(d-1)v'z'-zv'^{2}\partial_{z}f(v,z)\right]\label{eq:zpp}\\
 & -z\left[2z''+v'\left(2z'\partial_{z}f(v,z)+v'\partial_{v}f(v,z)\right)\right].\nonumber
\end{align}
 The extremal surface $\gamma_{\mathcal{A}}$ can be solved from (\ref{eq:vpp},\ref{eq:zpp})
as $v=\tilde{v}(x)$, $z=\tilde{z}(x)$. Note that the surface does
not live on a constant time slice for general $f(v,z)$.
Using the conserved Hamiltonian and the solution, we read the on-shell area
of the extremal surface $\gamma_{\mathcal{A}}$.
\begin{equation}\label{eq:areahee}
Area(\gamma_{\mathcal{A}})=2L^{d-2}\int_{0}^{l/2}\frac{z_{\ast}^{d-1}}{\tilde{z}(x)^{2d-2}}dx.
\end{equation}

\subsection{\label{subsec:SubregionComplexity}Subregion complexity }

Now we consider the extremal codimension-one hypersurface $\Gamma_{\mathcal{A}}$
enclosed by the extremal surface $\gamma_{\mathcal{A}}$ and $\mathcal{A}$.
We find that there are two equivalent ways to describe $\Gamma_{\mathcal{A}}$.
One way is to parametrizes $\Gamma_{\mathcal{A}}$ by $v(z)$ and the other
by $z(v)$. The parametrization $v(z)$ is more intuitive for the static backgrounds,
while the parametrization  $z(v)$ is more convenient for the dynamical backgrounds.

\subsubsection{Parametrization $v(z)$}

The bulk region enclosed by the extremal surface $v=\tilde{v}(x)$,
$z=\tilde{z}(x)$ can be parametrized by $v=v(z,x)$ generically. However,
due to the translational symmetry of the Vaidya metric (\ref{eq:VaidyaMetric}),
the parametrization which characterizes the extremal surface $\Gamma_{\mathcal{A}}$
should be independent of the coordinate $x$. Thus the extremal codimension-one hypersurface $\Gamma_{\mathcal{A}}$ can be parametrized by
\begin{equation}
v=v(z).
\end{equation}
The induced metric on $\Gamma_{\mathcal{A}}$ is
\begin{equation}
ds^{2}=\frac{1}{z^{2}}\left[-\left(f(v,z)\frac{\partial v}{\partial z}+2\right)\frac{\partial v}{\partial z}dz^{2}+dx^{2}+\sum_{i=1}^{d-2}dy_{i}^{2}\right].
\end{equation}
 The volume is
\begin{equation}
V=2L^{d-2}\int_{0}^{z_{\ast}}dz\int_{0}^{\tilde{x}(z)}dx\left[-f(v,z)\left(\frac{\partial v}{\partial z}\right)^{2}-2\frac{\partial v}{\partial z}\right]^{1/2}z^{-d}
\end{equation}
 where $\tilde{x}(z)$ is the codimension-two extremal surface $\gamma_{\mathcal{A}}$.
From the reduced Lagrangian
\begin{equation}
\mathcal{L}_{V}=\left[-f(v,z)\left(\frac{\partial v}{\partial z}\right)^{2}-2\frac{\partial v}{\partial z}\right]^{1/2}z^{-d},
\end{equation}
 one can read the equation of motion
\begin{equation}
0=v_{z}\left[4d+v_{z}\left[6df(v,z)-3z\partial_{z}f(v,z)+\left(2df(v,z)^{2}-zf(v,z)\partial_{z}f-z\partial_{v}f(v,z)\right)v_{z}\right]\right]+2zv_{zz}.\label{eq:VolumeEqVZ}
\end{equation}
This equation can be solved directly with the boundary condition
determined by $\gamma_{\mathcal{A}}=(\tilde{v}(x),\tilde{z}(x))$
and $\mathcal{A}$. However, there is a recipe for working out the
solution $\Gamma_{\mathcal{A}}$. In fact, we can get a relation $\tilde{v}(\tilde{z})$
from $\tilde{v}(x)$ and $\tilde{z}(x)$ by eliminating the parameter
$x$ on $\gamma_{\mathcal{A}}$. Then the extremal codimension-one
hypersurface $\Gamma_{\mathcal{A}}$ can be obtained by dragging $\tilde{v}(\tilde{z})$
along the $x$ direction. We have checked that $\tilde{v}(\tilde{z})$
is indeed the solution of (\ref{eq:VolumeEqVZ}). For all $x$
on $\Gamma_{\mathcal{A}}$ we have $\frac{\partial v}{\partial z}=\frac{\partial\tilde{v}}{\partial x}/\frac{\partial\tilde{z}}{\partial x}$.
So the on-shell volume is simply
\begin{equation}
V=2L^{d-2}\int_{0}^{z_{\ast}}dz\left[-f(v,z)\left(\frac{\partial\tilde{v}}{\partial x}/\frac{\partial\tilde{z}}{\partial x}\right)^{2}-2\frac{\partial\tilde{v}}{\partial x}/\frac{\partial\tilde{z}}{\partial x}\right]^{1/2}z^{-d}\tilde{x}(z).\label{eq:VolumeVZ}
\end{equation}
 This integral is more intuitive for the static background, as we
will show below. However, there are situations where $z_{\ast}$ is
a multi-valued function of boundary time $t$. In this case, $\tilde{v}(z)$
and $\tilde{x}(z)$ are also multi-valued functions of $z$. The integral
in (\ref{eq:VolumeVZ})  is then ill-defined. In these cases,  we choose another
parametrization to describe the extremal codimension-one hypersurface
$\Gamma_{\mathcal{A}}$.

\subsubsection{An alternative parametrization $z(v)$}

The extremal bulk region $\Gamma_{\mathcal{A}}$ enclosed by the extremal
surface $v=\tilde{v}(x)$, $z=\tilde{z}(x)$ can also be parametrized
by
\begin{equation}
z=z(v)
\end{equation}
 due to the translational symmetry of the Vaidya metric. The induced metric
on $\Gamma_{\mathcal{A}}$ now is
\begin{equation}
ds^{2}=\frac{1}{z^{2}}\left[-\left(f(v,z)+2z_v\right)dv^{2}+dx^{2}+\sum_{i=1}^{d-2}dy_{i}^{2}\right],
\end{equation}
 where $z_v=\partial z /\partial v$. The volume
\begin{equation}
V=2L^{d-2}\int_{v_{\ast}}^{\tilde{v}(l/2)}dv\int_{0}^{\tilde{x}(v)}dx\left[-f(v,z)-2z_v\right]^{1/2}z^{-d},
\end{equation}
 where $\tilde{x}(v)$ is the codimension-two extremal surface $\gamma_{\mathcal{A}}$.
The equation of motion gives
\begin{equation}
0=2d f(v,z)^2+4d z_{v}^2-3zz_v\partial_z f(v,z)+f(v,z)[6d z_v-z\partial_z f(v,z)]
-z[2z_{vv}+\partial_v f(v,z)].\label{eq:VolumeEqZV}
\end{equation}
The boundary condition is determined by the codimension-two surface
$\gamma_{\mathcal{A}}=(\tilde{v}(x),\tilde{z}(x))$ and $\mathcal{A}$.
Similar to the above subsection, the solution to Eq.(\ref{eq:VolumeEqZV})
can be determined by $\tilde{z}(\tilde{v})$ on the boundary $\gamma_{\mathcal{A}}$.
Then the on-shell volume reads
\begin{equation}
V=2L^{d-2}\int_{v_{\ast}}^{\tilde{v}(l/2)}dv\left[-f(v,z(v))-2\frac{\partial z}{\partial v}\right]^{1/2}z(v)^{-d}\tilde{x}(v).\label{eq:VolumeZV}
\end{equation}
 It  turns out that $\tilde{z}(\tilde{v})$ is a single-valued
function of $\tilde{v}$ all the time. Thus the integral in Eq.(\ref{eq:VolumeZV})
is well defined in the whole process of evolution. We will adopt this
formula to calculate the subregion complexity for the dynamical Vaidya-AdS spacetime.
Definitely, both Eq.(\ref{eq:VolumeVZ}) and (\ref{eq:VolumeZV})
give the same result when $\tilde{v}(\tilde{z})$ is singly valued.

\subsection{Static examples}

Since the Vaidya metric interpolates between the pure $AdS$ and the $SAdS$
black hole background, let us study the HEE
and the holographic subregion complexity in the pure $AdS$ and $SAdS$ backgrounds before we
discuss the
dynamical Vaidya background.

\subsubsection{Pure AdS}

For the pure AdS, we have $f(v,z)=1$.
The equations (\ref{eq:vpp},\ref{eq:zpp}) have a solution
\begin{equation}
v(x)=t-z(x).
\end{equation}
Here $t$ is the time coordinate on the boundary. Then Eq.(\ref{eq:RTt})
gives
\begin{equation}
\frac{dz}{dx}=\pm\sqrt{\frac{z_{\ast}^{2d-2}}{z^{2d-2}}-1}\label{eq:ZX}
\end{equation}
 where the plus sign is taken for $x<0$ and the minus sign for $x>0$.
Integrating the above formula gives a relation between $z_{\ast}$
and $l$.
\begin{equation}
z_{\ast}\frac{\sqrt{\pi}\Gamma(\frac{d}{2d-2})}{\Gamma(\frac{1}{2d-2})}=\frac{l}{2}.\label{eq:lengthzstar}
\end{equation}
 The on-shell area of the extremal surface reads
\begin{equation}
Area(\gamma_{\mathcal{A}})_{AdS_{d+1}}=\left(\frac{L}{\epsilon}\right)^{d-2}\frac{2}{d-2}-\left(\frac{L}{l}\right)^{d-2}\frac{2^{d-1}\pi^{\frac{d-1}{2}}}{(d-2)}\left(\frac{\Gamma(\frac{1}{2d-2})}{\Gamma(\frac{d}{2d-2})}\right)^{1-d}.
\end{equation}
 The result is the same as (37) in \cite{Albash2010}. The divergent
term is proportional to the area of the boundary of $\mathcal{A}$.
For $AdS_{3}$, we get
\begin{equation}
Area(\gamma_{\mathcal{A}})_{AdS_{3}}=2\log\left(\frac{l}{\epsilon}\right).
\end{equation}


The equation of motion (\ref{eq:VolumeEqVZ}) or (\ref{eq:VolumeEqZV})
for pure AdS can be solved directly as
\begin{equation}
v(z)=t-z.
\end{equation}
 Here $t$ is the time coordinate on the AdS boundary. The on-shell
volume reads
\begin{equation}
V=2L^{d-2}\int_{0}^{z_{\ast}}dzz^{-d}\int_{z}^{z_{\ast}}\left(\sqrt{\frac{z_{\ast}^{2d-2}}{\hat{z}^{2d-2}}-1}\right)^{-1}d\hat{z}
\end{equation}
 where we have used (\ref{eq:ZX}). Integrating it directly, we
get
\begin{equation}
V_{AdS_{d+1}}=\frac{L^{d-2}}{\epsilon^{d-1}}\frac{l}{d-1}+\frac{\sqrt{\pi}L^{d-2}}{z_{\ast}^{d-2}}\left(\frac{2\Gamma(\frac{d}{2d-2})}{\Gamma(\frac{1}{2d-2})}-\frac{d\Gamma(\frac{1}{2d-2})}{(d-1)^{2}\Gamma(\frac{d}{2d-2})}\right).\label{eq:AdSComplexity}
\end{equation}
 Note that the divergent term is proportional to the volume of $\mathcal{A}$.
From (\ref{eq:lengthzstar}, \ref{eq:AdSComplexity}), it is obvious
that the finite term has the same dependence of $l$ as the finite
part of the HEE.

\subsubsection{Schwarzschild-AdS black hole}

For the Schwarzschild-AdS black hole $f(v,z)=f(z)=1-mz^{d}$. The event
horizon locates at $z_{h}=m^{-1/d}$.
One can show that the solution to the equations (\ref{eq:vpp},\ref{eq:zpp})
is
\begin{equation}
v(x)=t+g(z(x)),\hspace{3ex} \partial_{z(x)}g(z(x))=-\frac{1}{f(z(x))}.
\end{equation}
 Here $t$ is the time coordinate on the AdS boundary. The conserved
Hamiltonian leads to a relation between $l$ and $z_{\ast}$.
\begin{equation}
\int_{\epsilon}^{z_{\ast}}\left[(1-mz^{d})\left(\frac{z_{\ast}^{2d-2}}{z^{2d-2}}-1\right)\right]^{-1/2}dz=\int_{0}^{l/2}dx=\frac{l}{2}.
\end{equation}
The on-shell area of the extremal surface turns out to be
\begin{equation}
Area(\gamma_{\mathcal{A}})_{SAdS_{d+1}}=2L^{d-2}\int_{z_{\ast}}^{\epsilon}\frac{z_{\ast}^{d-1}}{z^{2d-2}}\left[(1+mz^{d})\left(\frac{z_{\ast}^{2d-2}}{z^{2d-2}}-1\right)\right]^{-1/2}dz.
\end{equation}
 These two integrals have no explicitly analytical expression.


The equation of motion (\ref{eq:VolumeEqVZ}) about the codimension-one extremal surface $\Gamma_{\mathcal{A}}$ becomes
\begin{equation}
0=v_{z}\left[d\left(4-(2-mz^{d})v_{z}(-3-(1-mz^{d})v_{z})\right)\right]+2zv_{zz}.
\end{equation}
 One can verify easily that the solution is
\begin{equation}
v=t+g(z),\hspace{3ex}\partial_{z}g(z)=\frac{1}{-f(z)},
\end{equation}
and read the
on-shell volume of $\Gamma_{\mathcal{A}}$
\begin{equation}
V_{SAdS_{d+1}}=2L^{d-2}\int_{0}^{z_{\ast}}dz\frac{1}{\sqrt{1-mz^{d}}}z^{-d}\int_{z}^{z_{\ast}}\left(\sqrt{\frac{z_{\ast}^{2d-2}}{s^{2d-2}}-1}\right)^{-1}ds.
\end{equation}
 This is the same as (2.7) in \cite{subBenAmi2016}.

\section{Subregion complexity in Vaidya-AdS spacetime}

We study the evolution of the subregion complexity after a thermal quench in this section. The thermal quench in CFT could be described holographically by the dynamical Vaidya spacetime, whose initial state corresponds to the pure AdS and the final state corresponds to the SAdS black hole.  As we have done for the static cases in the previous subsection, we need first work out the evolution of the codimension-two extremal surface $\gamma_A$.

\subsection{Evolution of holographic entanglement entropy}

For the Vaidya metric (\ref{eq:VaidyaMetric}) with the mass function (\ref{eq:Massfunction}),
the equations (\ref{eq:vpp},\ref{eq:zpp}) for the HEE
become
\begin{align}
0= & 2-2d-\frac{1}{2}\left[4-4d+(d-2)M\left(1+\tanh(\frac{v}{v_{0}})\right)z^{d}\right]v'^{2}+4(d-1)v'z'+2zv''\\
0= & (d-2)M^{2}\left(1+\tanh(\frac{v}{v_{0}})\right)^{2}z^{2d}v'^{2}+2\frac{Mz^{d+1}v'^{2}}{v_{0}\cosh^{2}(v/v_{0})}+8(d-1)(-1+v'^{2}+2v'z'),\nonumber \\
 & -2M\left(1+\tanh(\frac{v}{v_{0}})\right)z^{d}(2-2d+(3d-4)v'^{2}+2(d-2)v'z')-8zz''.
\end{align}
We solve these two equations numerically by using the shooting method
with the boundary conditions,
\begin{equation}
v'(0)=z'(0)=0,\ \ z(0)=z_{\ast},\ v(0)=v_{\ast}.
\end{equation}
Here $(v_{\ast},z_{\ast})$ is the turning point of the extremal surface
$\gamma_{\mathcal{A}}$. The targets on the AdS boundary are
\begin{equation}
z(l/2)=\epsilon,\ \ v(l/2)=t-\epsilon
\end{equation}
 where $\epsilon$ is a cutoff and $t$ is the boundary time.

\begin{figure}
\begin{centering}
\begin{tabular}{cc}
\includegraphics[scale=0.65]{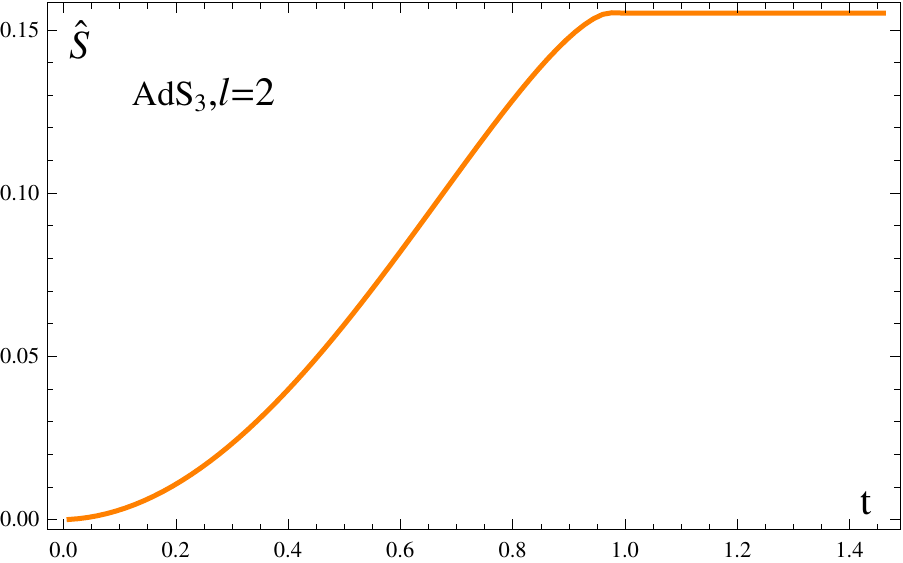} & \includegraphics[scale=0.65]{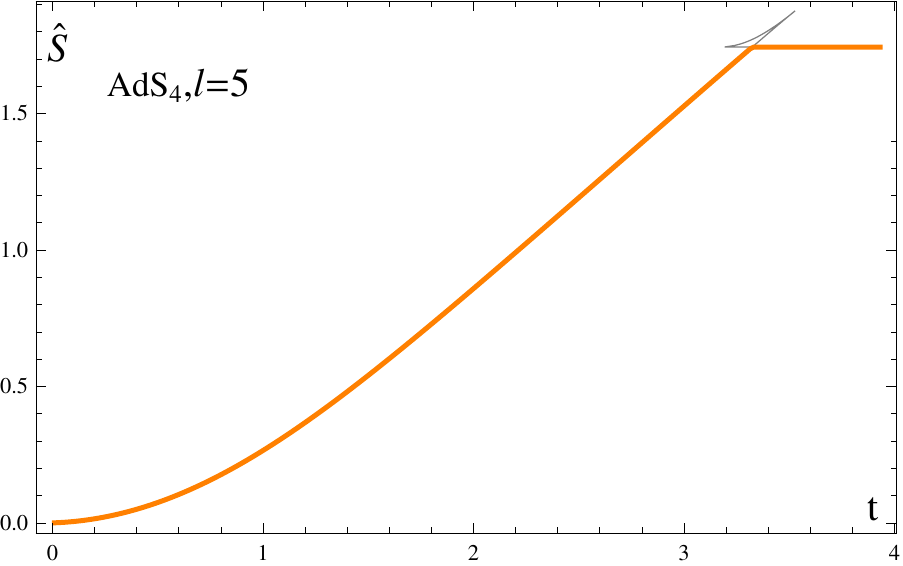}\tabularnewline
\end{tabular}
\par\end{centering}
\caption{\label{fig:EEevolution}The evolution of holographic entanglement
entropy with respect to the boundary time $t$. We fix $M=1$ and
$v_{0}=0.01$ here. The transition point in the right panel locates
at $t=3.3248,\hat{S}=1.7437$. }
\end{figure}

 Once we get the solution, the HEE can be obtained from (\ref{eq:areahee}). As the HEE is divergent,
 it is convenient to define  subtracted HEE
 \begin{equation}
 \hat S_{HEE}=S_{HEE,Vaidya}-S_{HEE,AdS},
 \end{equation}
 where both $S_{HEE,Vaidya}, S_{HEE,AdS}$ are defined with respect to the same boundary region. As we are discussing the strip which has a finite width but infinite length, we furthermore define a finite quantity from the subtracted HEE
 \begin{equation}
\hat{S}=\frac{4G_{N}\hat S_{HEE}}{2L^{d-2}}=\frac{Area(\gamma_{A})_{Vaidya}-Area(\gamma_{A})_{AdS_{d+1}}}{2L^{d-2}}.
\end{equation}
Its evolution has two typical profiles as shown in Fig. \ref{fig:EEevolution}. The first profile appears in the AdS$_3$ case and also in the higher $AdS_{d+1} (d\geq 3)$ cases with narrow strips. In these cases, the HEE in the Vaidya-AdS spacetime
 increase
monotonically and reach saturation at late time. In the left panel of Fig. \ref{fig:EEevolution}, we show the evolution of $\hat S$ for an interval of length $l=2$ in $AdS_{3}$. The other profile appears in the higher $AdS_{d+1} (d\geq 3)$ cases with  wide strips. We show this profile in the right panel of Fig. \ref{fig:EEevolution}, which corresponds to the
 strip of width $l=5$ in $AdS_{4}$. Different from the first profile,  this profile shows that though the HEE increases first as well,  it exhibits a swallow tail before reaching the saturation. This phenomenon was first discovered in \cite{Albash2010}.
The swallow tail implies that there are
multiple solutions to the differential equationns at a given boundary time. We should choose the one which gives the surface of the minimum area. The solutions which correspond to the surfaces of non-minimum  area are marked in grey in the right panel of Fig. \ref{fig:EEevolution}. In any case, the HEE is always increasing continuously before reaching the saturation. For more details on the evolution of the HEE after a thermal quench, please refer to \cite{Albash2010,Balasubramanian1103,Liu1311,Li2013}.

\begin{figure}
\begin{centering}
\begin{tabular}{cc}
\includegraphics[scale=0.7]{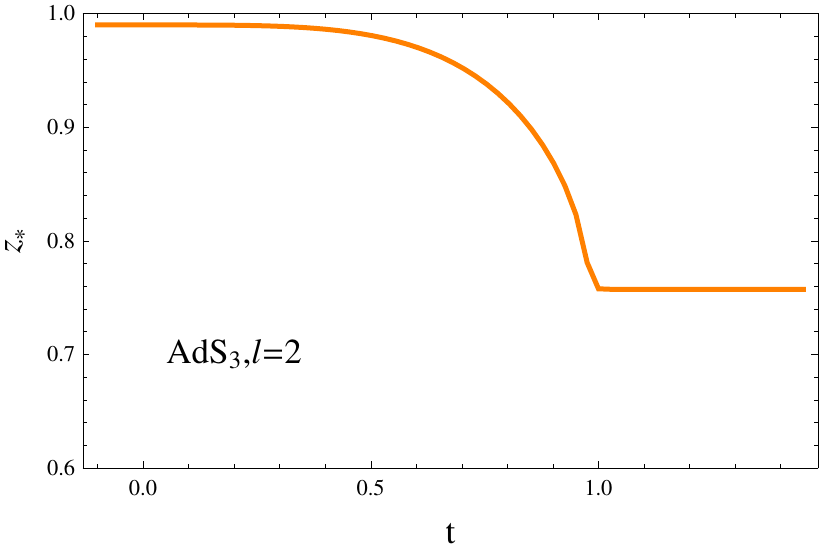} & \includegraphics[scale=0.7]{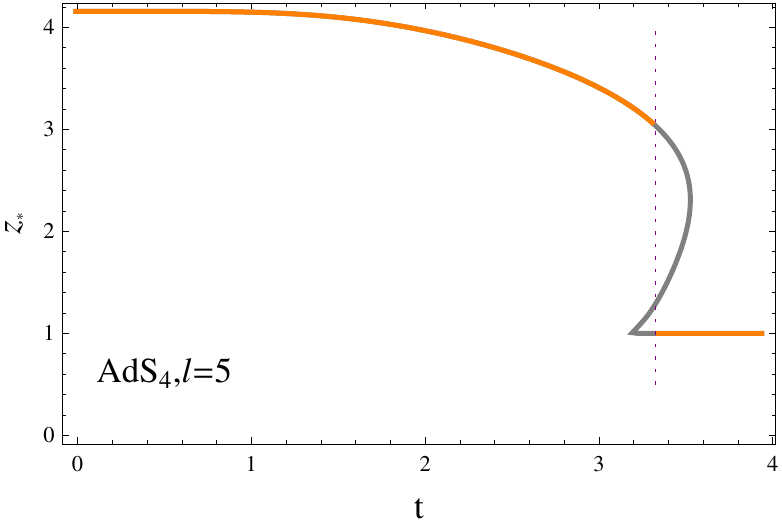}\tabularnewline
\end{tabular}
\par\end{centering}
\caption{\label{fig:ZsEvolution}The evolution of $z_{\ast}$ with respect
to boundary time $t$. We fix $M=1$ and $v_{0}=0.01$ here.
The orange parts correspond to the surfaces of the minimum  area. The transition point locates
at $t=3.3248$ in the right panel. }
\end{figure}

 In Fig. \ref{fig:ZsEvolution} we show the corresponding evolution of $z_*$.  It is multi-valued only when $\hat{S}$ is multi-valued. The multi-valuedness depends on the spacetime dimension and the strip width.  In $AdS_3$, $z_*$ is always singly valued no matter how large $l$ is. However, in the spacetime with dimension $d\geq 4$,  $z_*$ is singly valued only when $l$ is small. When $l$ is large enough, $z_*$ becomes multi-valued. For the  $AdS_4$  we study here, the critical width is $l=1.6$. When $z_*$ is multi-valued, its evolution is subtle. The multi-valuedness means that there are multiple extremal surfaces at a given time. The requirement\cite{HRT1509} that the HRT surface  should be of the minimal area leads to the transition at some point. In the right panel of Fig. \ref{fig:ZsEvolution}, the evolution of $z_*$ follows the line in orange, which has discontinuity. The transition point is at $t=3.3248$.

\begin{figure}[h]
\begin{centering}
\begin{tabular}{cc}
\includegraphics[scale=0.7]{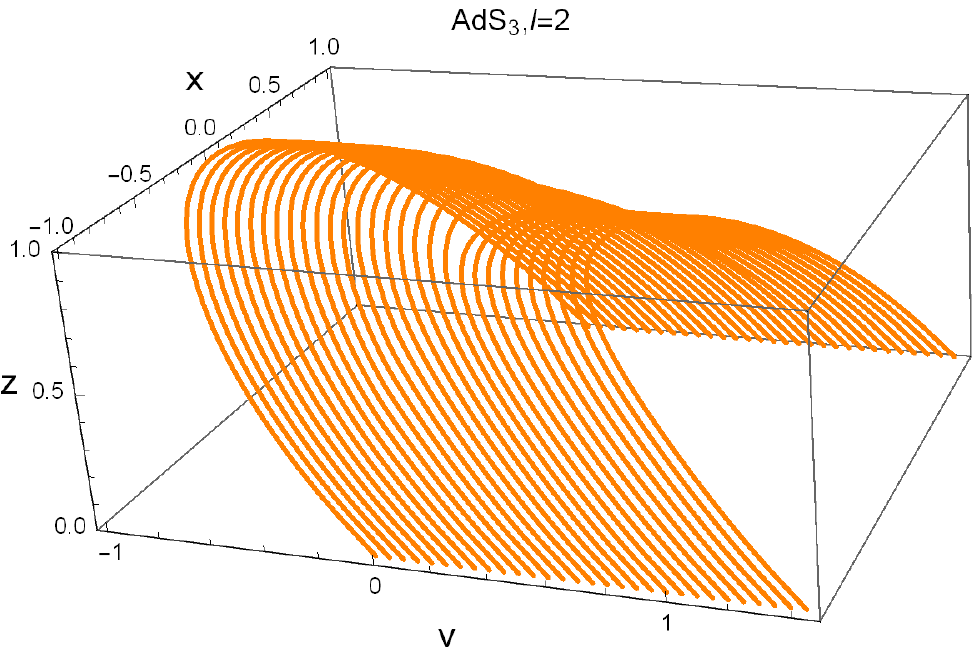} & \includegraphics[scale=0.7]{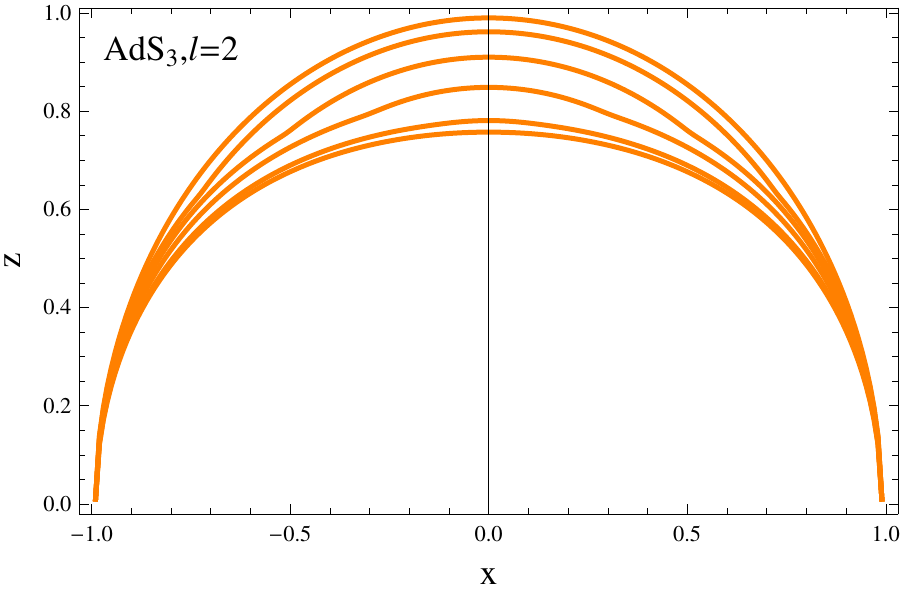}\tabularnewline
\end{tabular}
\par\end{centering}
\caption{\label{fig:ExtremalSurfaceProfile}The evolution of extremal surface
$\gamma_{\mathcal{A}}=(\tilde{z}(x),\tilde{v}(x))$ for $AdS_{3}$
and $l=2$. We fix $M=1$ and $v_{0}=0.01$ here. The left panel shows
the evolution in $(x,v,z)$. The right panel shows their projection
on to the $(x,z)$ plane. The extremal surface evolves from left to
right in the left panel and from up to down in the right panel. }
\end{figure}

\begin{figure}[h]
\begin{centering}
\begin{tabular}{cc}
\includegraphics[scale=0.7]{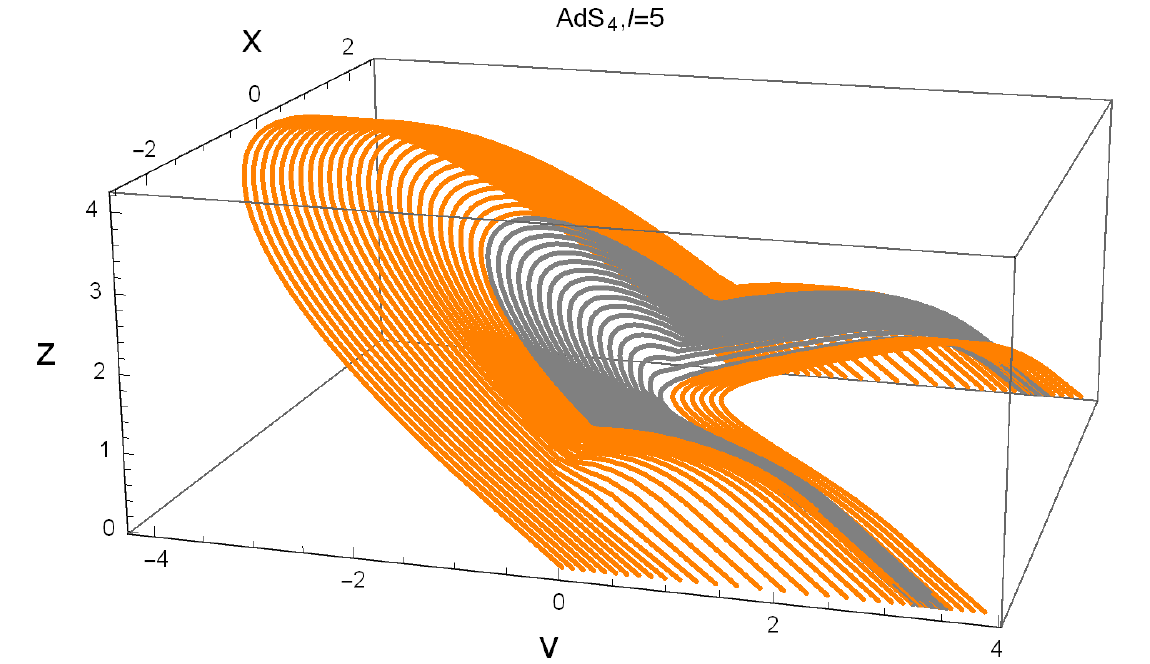} & \includegraphics[scale=0.7]{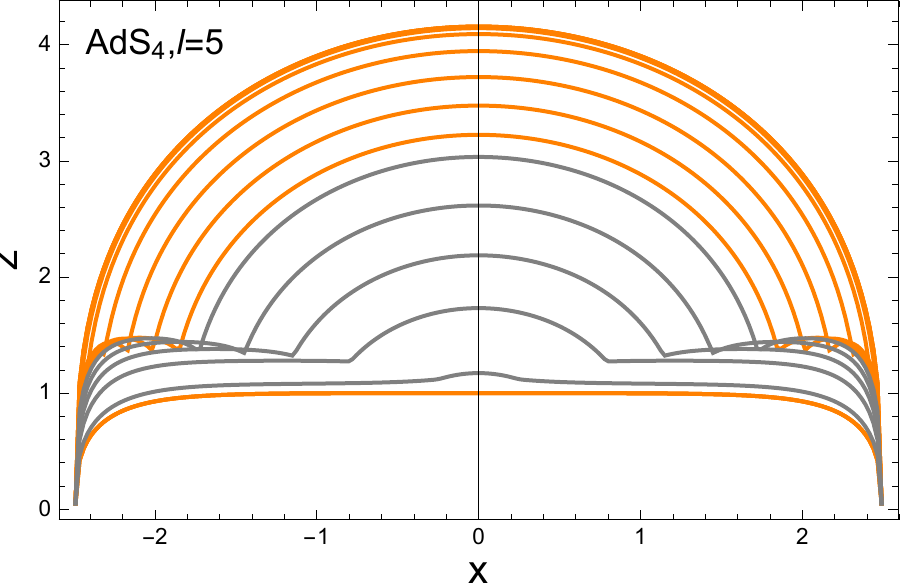}\tabularnewline
\end{tabular}
\par\end{centering}
\caption{\label{fig:ExtremalSurfaceProfile2}The evolution of extremal surface
$\gamma_{\mathcal{A}}=(\tilde{z}(x),\tilde{v}(x))$ for $AdS_{4}$
and $l=5$. We fix $M=1$ and $v_{0}=0.01$ here.  }
\end{figure}

 The details of the corresponding evolution of the extremal surface $\gamma_{\mathcal{A}}$ are shown in Fig. \ref{fig:ExtremalSurfaceProfile} and Fig. \ref{fig:ExtremalSurfaceProfile2}.
In Fig. \ref{fig:ExtremalSurfaceProfile},
$\gamma_{\mathcal{A}}$  evolves smoothly from the initial state to the
final state. In
Fig. \ref{fig:ExtremalSurfaceProfile2}, the evolution of $\gamma_{\mathcal{A}}$ has a gap marked in gray  before it reaches the final
state. These gray surfaces correspond to the swallow tail in Fig. \ref{fig:EEevolution}. They are not the smallest area surfaces at the given boundary times.


More precisely, the multi-valuedness not only depends on the spacetime dimension and the size of the strip, but also depends on the parameter $v_0$. In the above discussion, we fix $M=1$ and $v_0=0.01$. As we will show later, the swallow tail would disappear if we choose a large enough $v_0$, which corresponds to a  slow quench.

\subsection{Evolution of subregion complexity}

\begin{figure}
\begin{centering}
\includegraphics[scale=0.7]{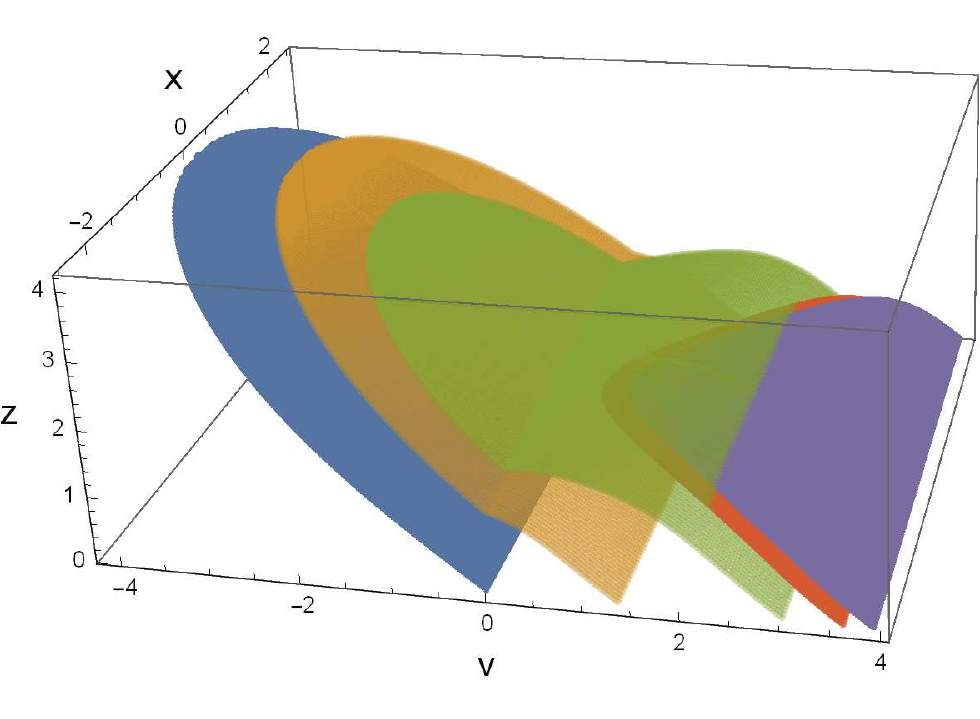}
\par\end{centering}
\caption{\label{fig:ComplexityOneSlice}The evolution of the codimension-one
extremal surface $\Gamma_{\mathcal{A}}$ which characterizes the subregion
complexity enclosed by the codimension-two
extremal surface $\gamma_{\mathcal{A}}$
and $\mathcal{A}$. We take $AdS_{4}, l=5,M=1$ and $v_{0}=0.01$ here.}
\end{figure}

Once we get the HRT surface $\gamma_{\mathcal{A}}=(\tilde{v}(x),\tilde{z}(x))$,
we can determine the codimension-one extremal surface $\Gamma_{\mathcal{A}}$
by dragging the points on $\gamma_{\mathcal{A}}$ along the $x$ direction,
as we have stressed in the subsection \ref{subsec:SubregionComplexity}.
The evolution of $\Gamma_{\mathcal{A}}$ has the profile shown in
Fig. \ref{fig:ComplexityOneSlice}. Similar to the HEE, the volume of $\Gamma_{\mathcal{A}}$
which can be obtained by (\ref{eq:VolumeVZ}) is divergent, thus we define a normalized subtracted volume \begin{equation}
\hat{C}=\frac{8\pi RGC_{\mathcal{A}}}{2L^{d-2}}=\frac{V_{Vaidya}-V_{AdS}}{2L^{d-2}}
\end{equation}
 where $R$ is the AdS radius which has been set to $1$, and the volumes are defined with respect to the same boundary region. It is finite and can be used to characterize the evolution of the subregion complexity.

\begin{figure}
\begin{centering}
\begin{tabular}{cc}
\includegraphics[scale=0.7]{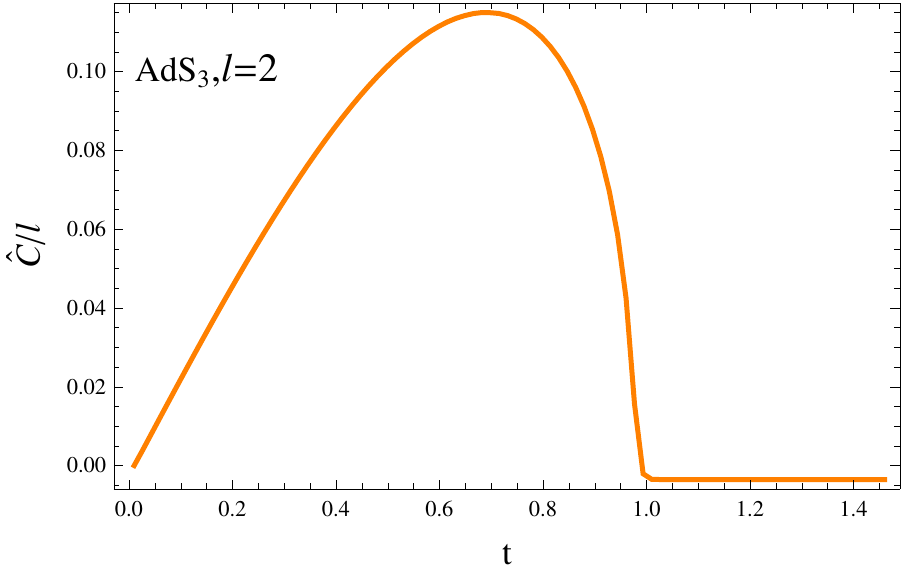} & \includegraphics[scale=0.7]{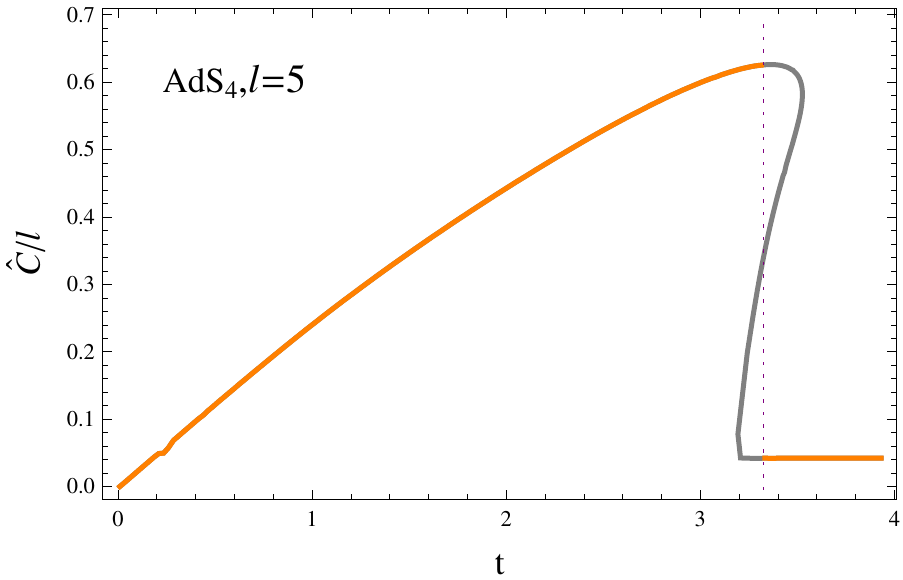}\tabularnewline
\end{tabular}
\par\end{centering}
\caption{\label{fig:ComplexityEvolution}The evolution of the subregion complexity
density $\hat{C}/l$ with respect to the boundary time $t$. We fix $M=1$ and
$v_{0}=0.01$ here. The transition point in the right panel locates at $t=3.3248$. }
\end{figure}

As shown in Fig. \ref{fig:ComplexityEvolution},
the evolution of the subregion complexity has a common feature: it increases at the early stage and reaches a maximum,   then it decreases and gets to saturation in the late time.

Another important feature  of the subregion complexity under a global quench is that it may evolves discontinuously,
as shown by the orange line in the right panel of Fig. \ref{fig:ComplexityEvolution}. This is due to the transition of the HRT surface shown in Fig. \ref{fig:ZsEvolution}.  As a result, the subregion complexity exhibits a sudden drop in the evolution.
The gray dashed part in  Fig. \ref{fig:ComplexityEvolution} corresponds to the swallow tail in Fig. \ref{fig:EEevolution}.
In other words, even though the HEE  always evolves continuously, the subregion complexity
does not.


\begin{figure}[H]
\begin{centering}
\begin{tabular}{cc}
\includegraphics[scale=0.7]{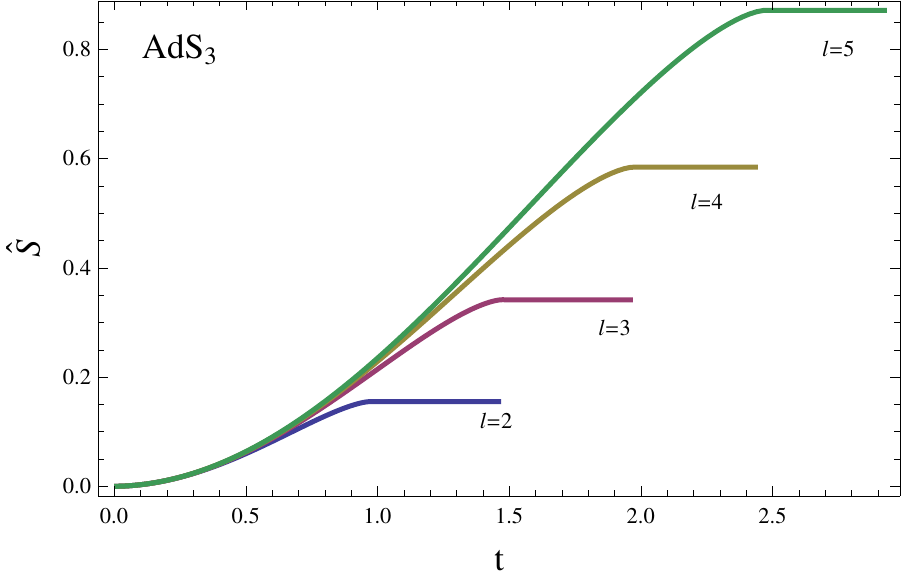} & \includegraphics[scale=0.7]{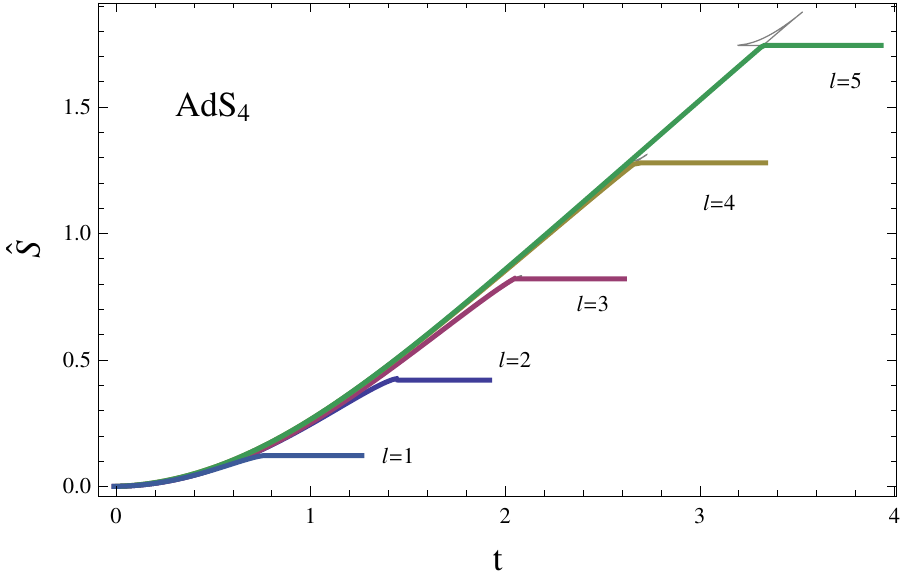}\tabularnewline
\includegraphics[scale=0.7]{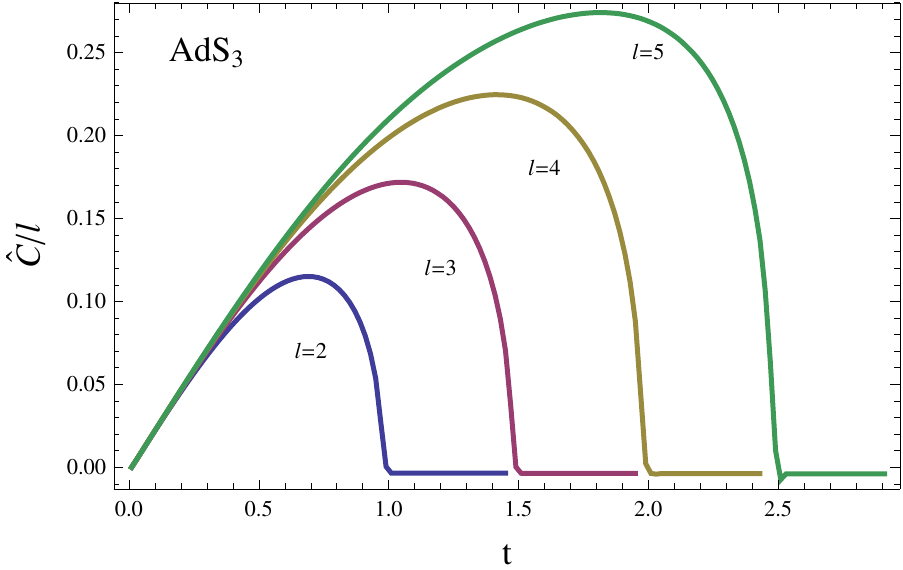} & \includegraphics[scale=0.7]{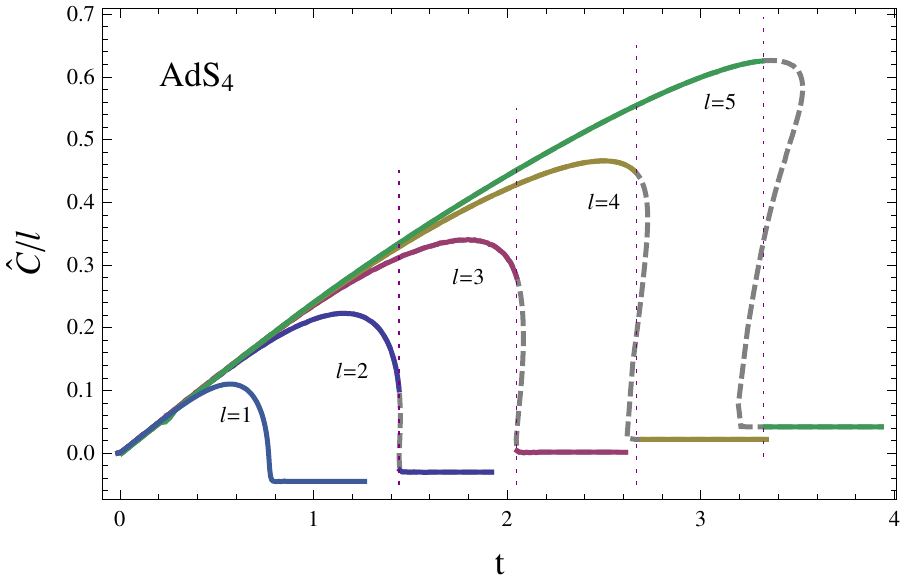}\tabularnewline
\end{tabular}
\par\end{centering}
\caption{\label{fig:ComplexityEntropyT}The upper panels show the evolution of the entanglement
entropy for different $l$.  The lower panels show the corresponding subregion
complexity density $\hat{C}/l$ (thick lines) for different $l$. We fix $M=1,v_{0}=0.01$ here.}
\end{figure}

\subsubsection{The dependence of subregion complexity evolution on $l$}

\begin{figure}
\begin{centering}
\begin{tabular}{c}
\includegraphics[scale=0.7]{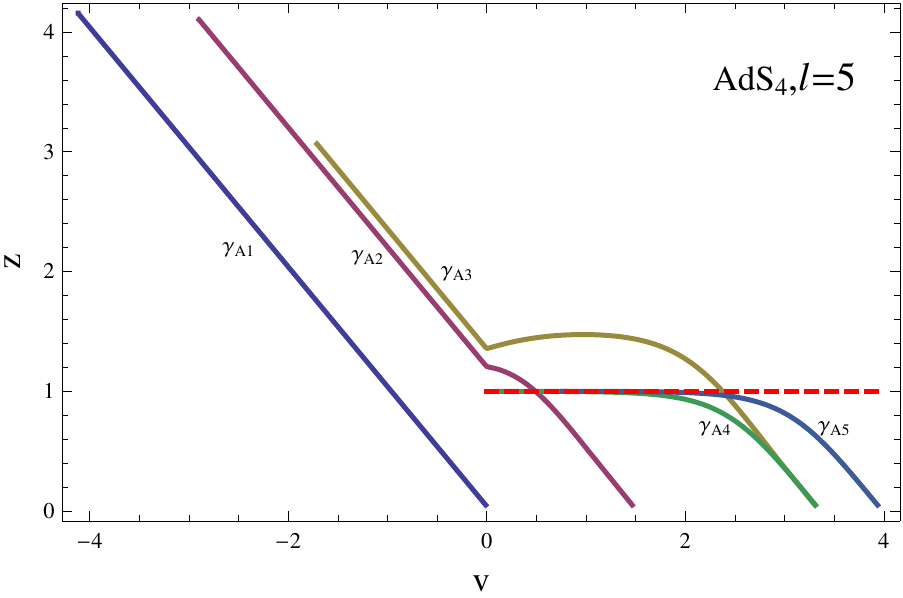}\tabularnewline
\end{tabular}
\par\end{centering}
\caption{\label{fig:Transition}The snaps of the evolution of entanglement
surface $\gamma_{\mathcal{A}}$ in the $(v,z)$ plane. The time flows
from left to right. $\gamma_{\mathcal{A}i} (i=1,2,3,4,5)$ are the entanglement surfaces corresponding to the boundary times $t=-0.01,1.4660,3.3248,3.3248$ and $3.9373$, respectively. $\gamma_{\mathcal{A}3}$ and $\gamma_{\mathcal{A}4}$
have the same area at boundary time
   $t=3.3248$,
 which corresponds
to the transition point. The red dashed line is the apparent horizon.
We fix $M=1,v_{0}=0.01$ here.}
\end{figure}

The evolutions of the holographic entanglement entropy and the subregion complexity  for different $l$  are displayed in Fig. \ref{fig:ComplexityEntropyT}. As shown in the lower left panel  for the Vaidya-$AdS_{3}$ spacetime, the subregion complexity increases at the early stage
and then decreases and maintains to be a constant value at late time. The situation in the Vaidya-$AdS_{4}$ spacetime is similar  except that when the size $l$ is large enough, there is a sudden drop of the subregion complexity in the evolution, as shown in the lower right panel. This corresponds exactly to the kink in the evolution of HEE shown in the upper right panel.
We plot the transition point in Fig. \ref{fig:Transition}. The entanglement
surface evolves from left to right. Its profile experiences a transition
at time $t=3.3248$. The corresponding surfaces $\gamma_{\mathcal{A}3}$
and $\gamma_{\mathcal{A}4}$ have the same area. But the volumes they
enclosed are different. This leads to a sudden drop of the subregion complexity.

Remarkably, we find that the growth rate of the complexity density
for different $l$ is almost the same at the early stage. This is very
similar to the evolution of the entanglement entropy for different $l$.
It has been argued that for the geometry of strip, the area of the
boundary of the subregion $\mathcal{A}$ does not change, so the initial
propagation of excitation from the subregion $A$ to outside which contributes
to the entanglement is not affected by the strip width \cite{Albash2010}.
Since  in the early time the complexity density growth  is mainly caused by the local excitations, which is
independent of $l$, the same  rate of increasing for different $l$ could
be expected. On the other hand, the nonlocal excitations have important contributions to the subregion complexity at later time such that the evolutions present different behaviors. 

\begin{figure}[H]
\begin{centering}
\begin{tabular}{cc}
\includegraphics[scale=0.7]{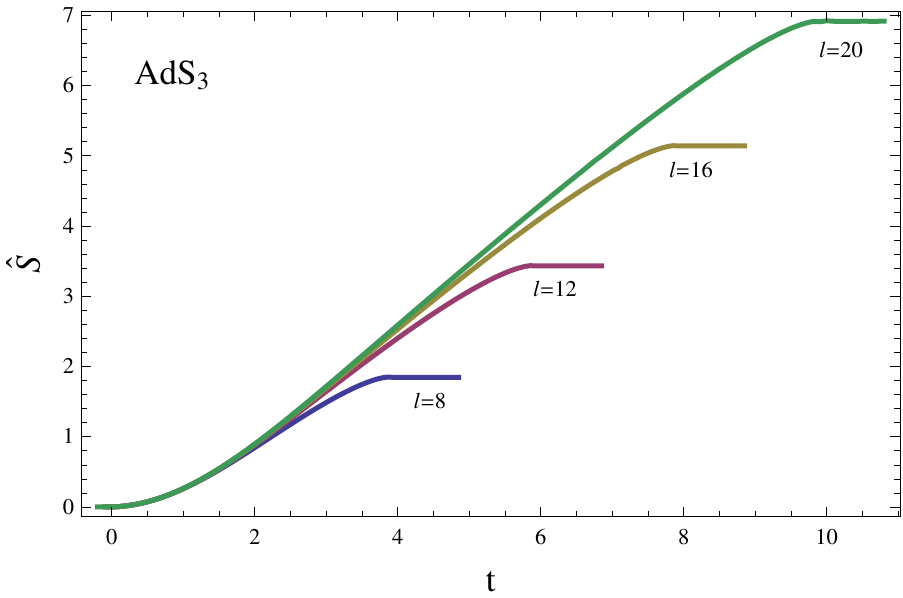} & \includegraphics[scale=0.7]{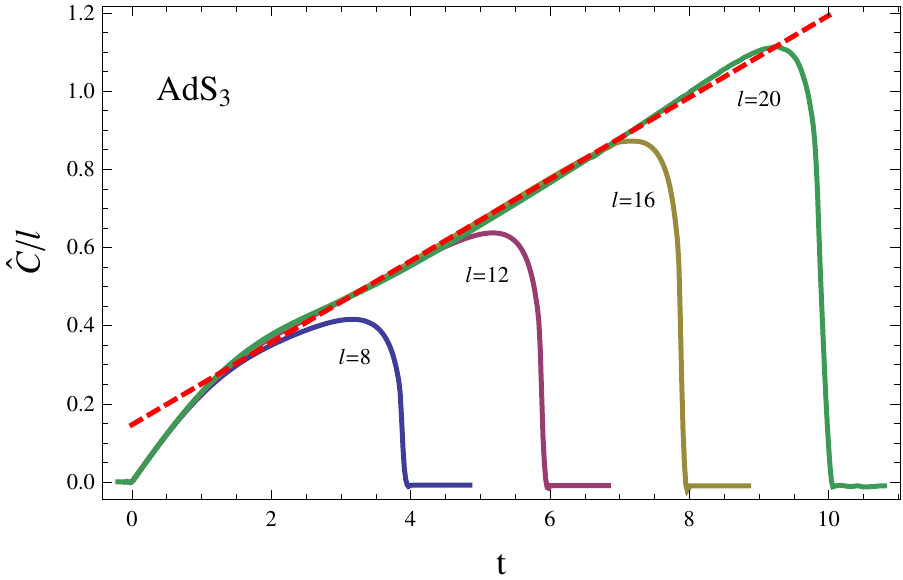}\tabularnewline
\end{tabular}
\par\end{centering}
\caption{\label{fig:ComplexityLlong} The evolution of the entanglement entropy
and the subregion complexity for different $l$ in $AdS_{3}$. We fix $M=1,v_{0}=0.01$ here.}
\end{figure}

For the cases that $l$ is large enough, we find that the complexity
density grows for a long time before it drops down. The evolutions of the subregion complexities for different $l$ in $AdS_{3}$ are shown in the right panel of Fig. \ref{fig:ComplexityLlong}. 
The complexity presents two increasing stages: it increases
faster in the early time, then it increases at a slower rate. At the second stage, it evolves almost linearly, the larger $l$ is, the longer it stands, with the slope being proportional to the mass parameter,
\begin{equation}
\hat{C}/l\propto Mt.
\end{equation}
Besides, we also notice that, the maximum value of the complexity density  in the evolution is proportional to the size $l$
\begin{equation}
\hat{C}_{max}/l\propto l.
\end{equation}
The proportional factor  is a function of spacetime dimension $d$ and the mass parameter $M$.

Due to the limitation of our numerical
method, the more detailed analysis on the evolution of the complexity for different $l$ in $AdS_{4}$ is
absent here. Nevertheless, from the right panel of Fig. \ref{fig:ComplexityEntropyT}, we see that the linear growth in the second stage persists, and the larger the size, the longer the complexity increases.

In fact, the linear growth of the complexity has been found in many different
non-holographic systems \cite{Susskind2014,pHashimoto2017} and also
appears in the CV and CA conjectures at late time limit \cite{Carmi2017,Kim2017}.
It is also reminiscent of the time evolution of the entanglement entropy
from black hole interiors \cite{Hartman2013}. In our model, if we
set $l\to\infty$, we may expect that the behavior of the complexity will turn
to the behavior for whole boundary region and so the complexity would increase
linearly at late time as well. Since the brutal numerical method is not able to
study the cases of an extremal large $l$, one may turn to the analytical way adopted in \cite{Liu1311}
  to study the linear growth
of the subregion complexity. Actually, { the recent studies of the complexity following a global quench based on the CA and CV conjectures show that the late time behavior of the complexity for the whole boundary region  is linear \cite{Moosa:2017yvt,Chapman:2018}.}

\subsubsection{The dependence of subregion complexity evolution on $v_{0}$}

\begin{figure}
\begin{centering}
\begin{tabular}{cc}
\includegraphics[scale=0.5]{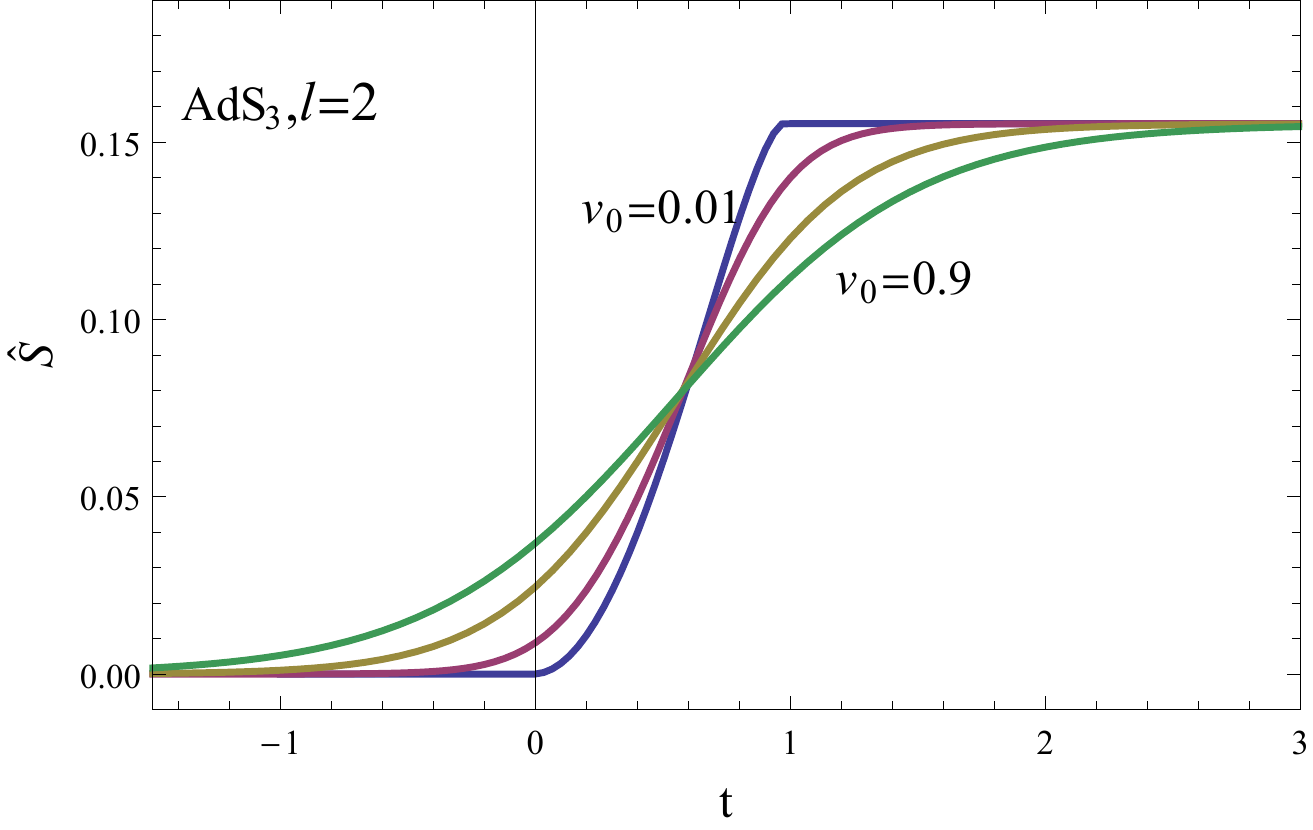} & \includegraphics[scale=0.5]{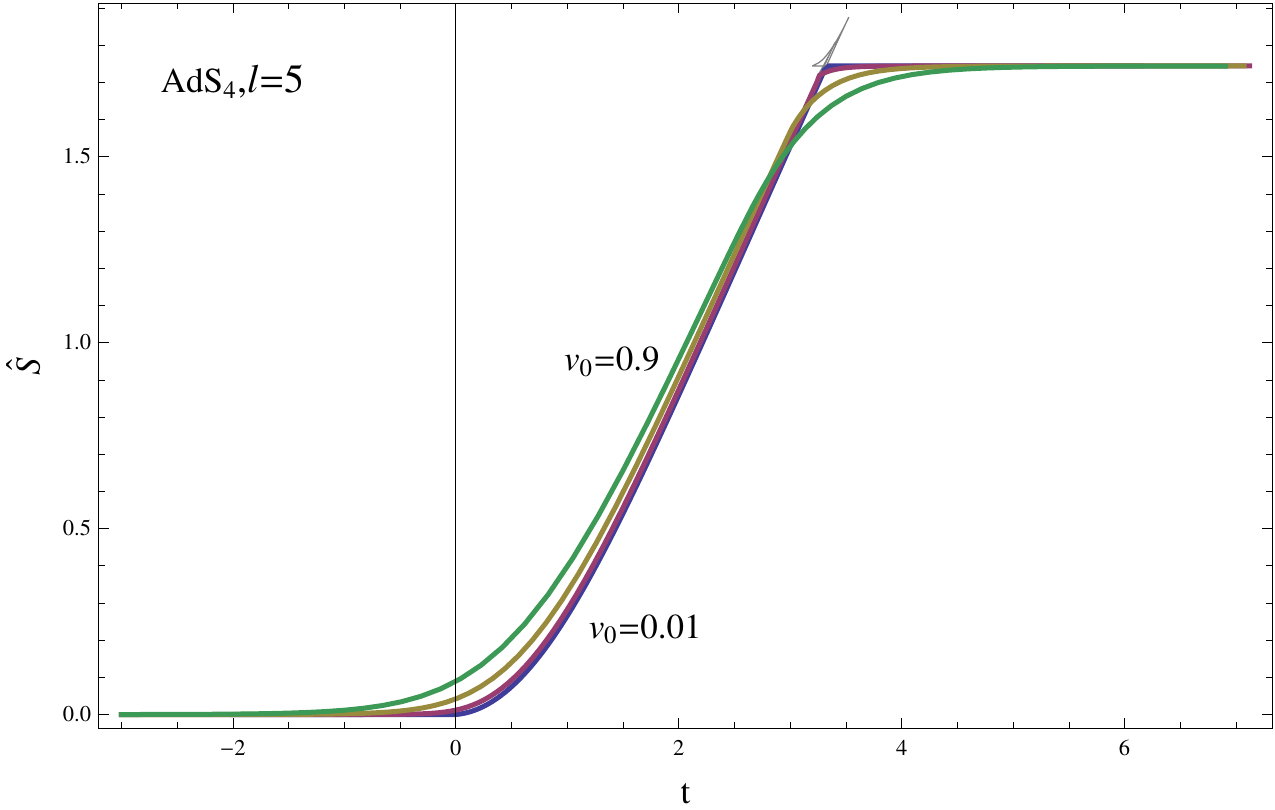}\tabularnewline
\includegraphics[scale=0.5]{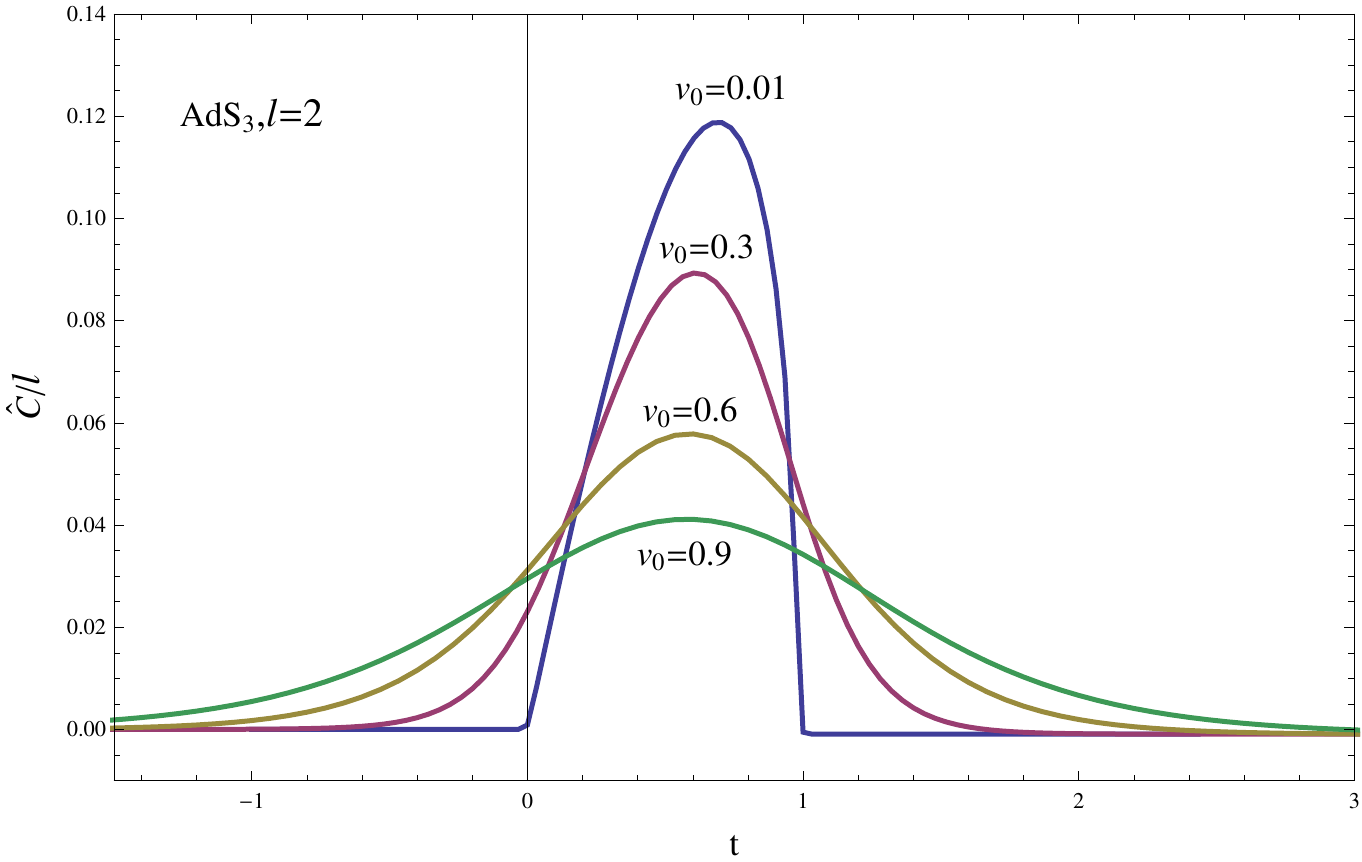} & \includegraphics[scale=0.56]{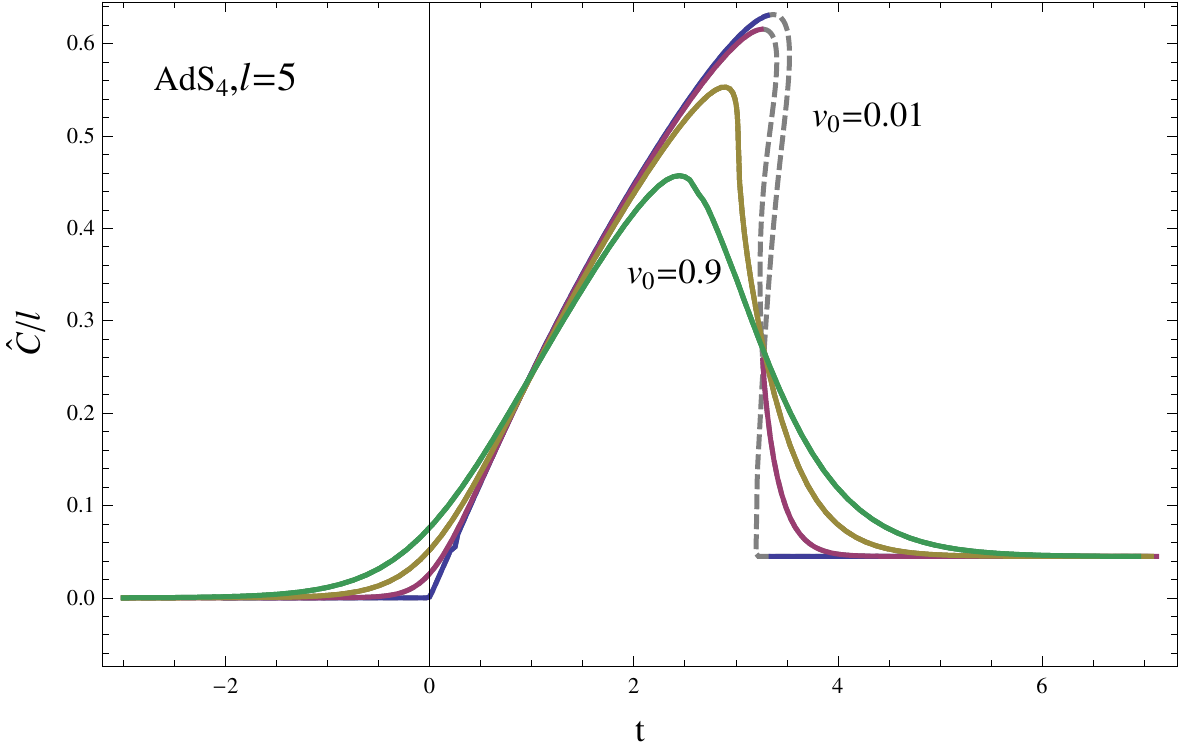}\tabularnewline
\end{tabular}
\par\end{centering}
\caption{\label{fig:ComplexityV0}The dependence of the entanglement entropy and
the subregion complexity density $\hat{C}/l$ on $v_{0}$. We
take $v_{0}=0.01,0.3,0.6,0.9$ and fix $M=1$ here. The sudden drop in the subregion complexity evolution disappears when $v_0>0.57$.}
\end{figure}

In this subsection, we study the effect of the parameter $v_{0}$ on the evolution of the subregion complexity. The parameter $v_0$  characterizes the thickness of the null-dust shell in the gravity, its inverse could be taken as the speed of the quench. The numerical results are shown in Fig. \ref{fig:ComplexityV0}. All
the processes evolve from the pure AdS background to an identical SAdS
black hole background. 
It is obvious that the thinner
the shell is, the sooner the quench happens, and the earlier the system reaches equilibrium.  The thicker
the shell is, the earlier the system starts to evolve,  but the maximum
complexity the system can reach  is smaller. Thus the subregion complexity is closely related to  the change rate of a state. Especially,   the sudden drop in the complexity evolution disappears when $v_0$ is large enough. For the AdS$_4$ case, the critical point is $v_0=0.57$. Namely, if the quench happens slowly enough, the subregion complexity evolves continuously.

\subsubsection{The dependence of subregion complexity evolution on $M$}

\begin{figure}
\begin{centering}
\begin{tabular}{cc}
\includegraphics[scale=0.5]{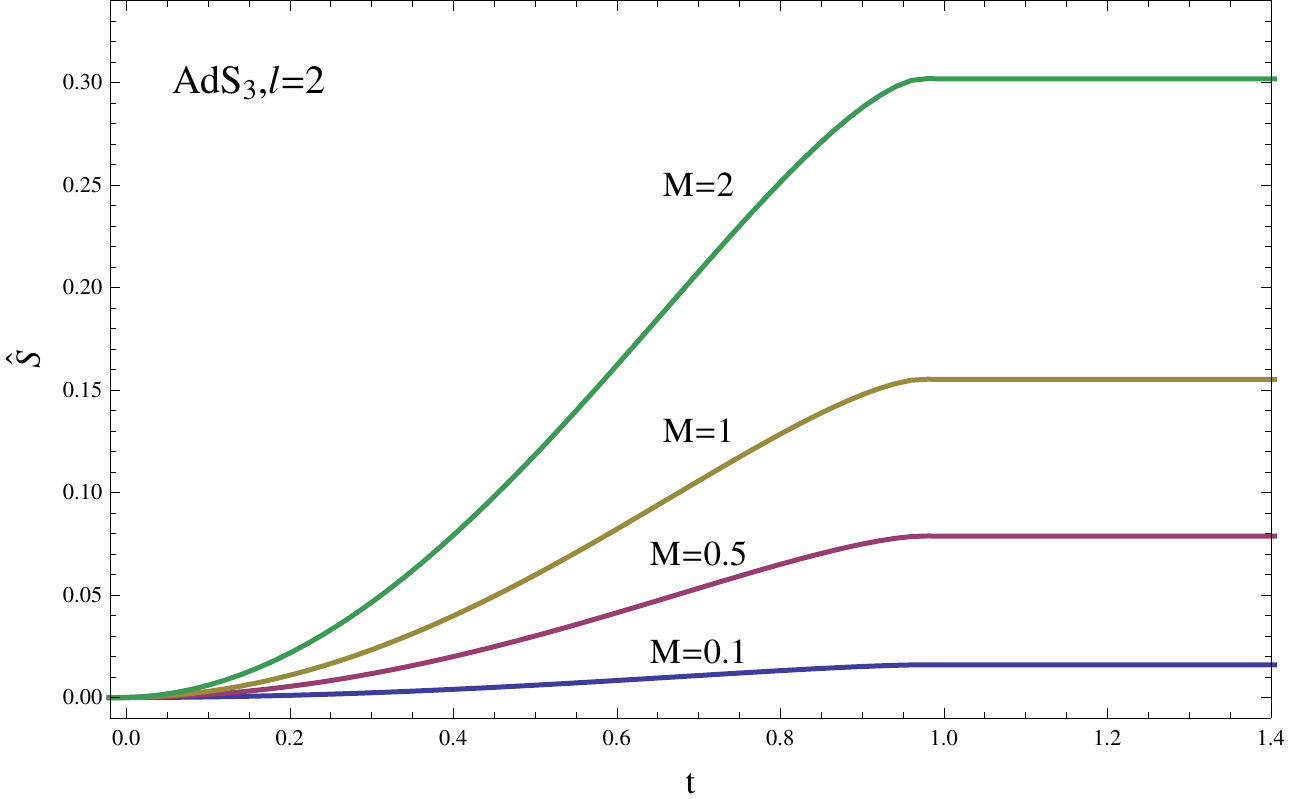} & \includegraphics[scale=0.56]{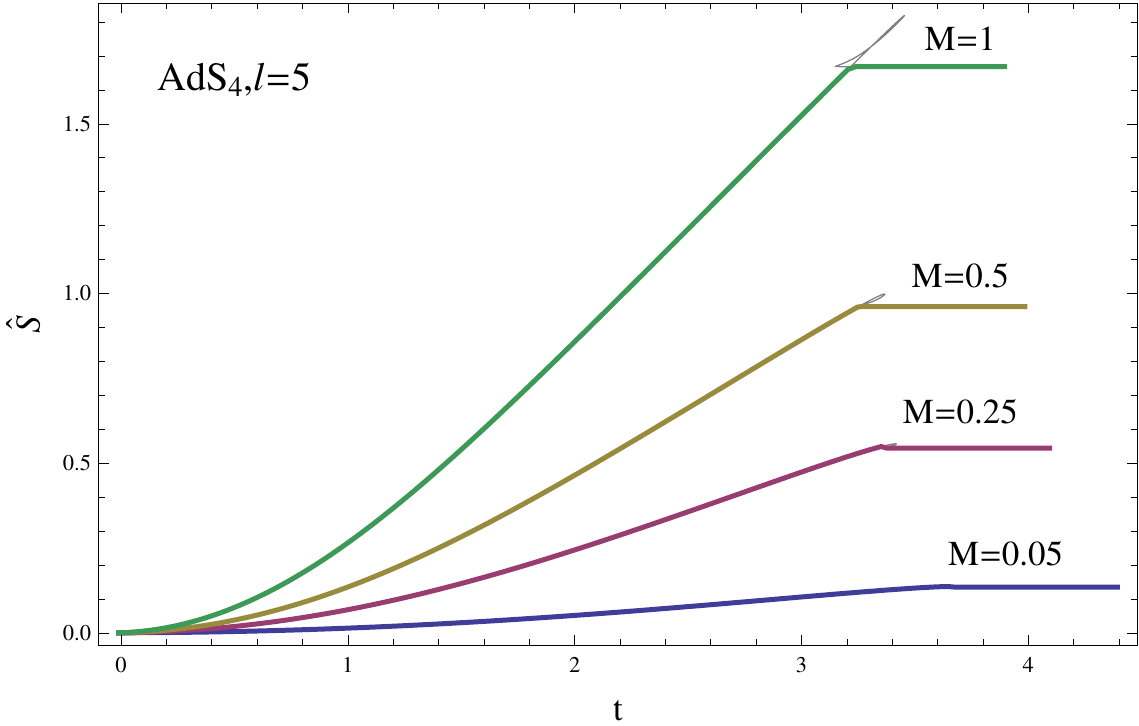}\tabularnewline
\includegraphics[scale=0.56]{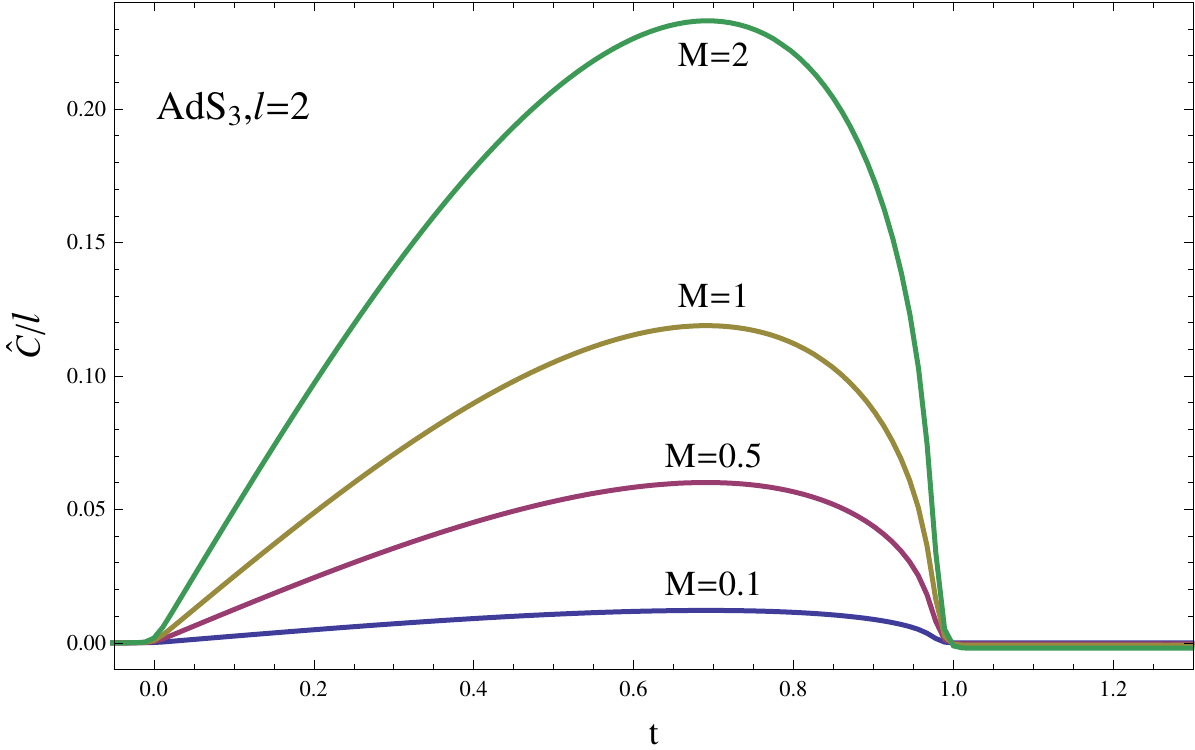} & \includegraphics[scale=0.6]{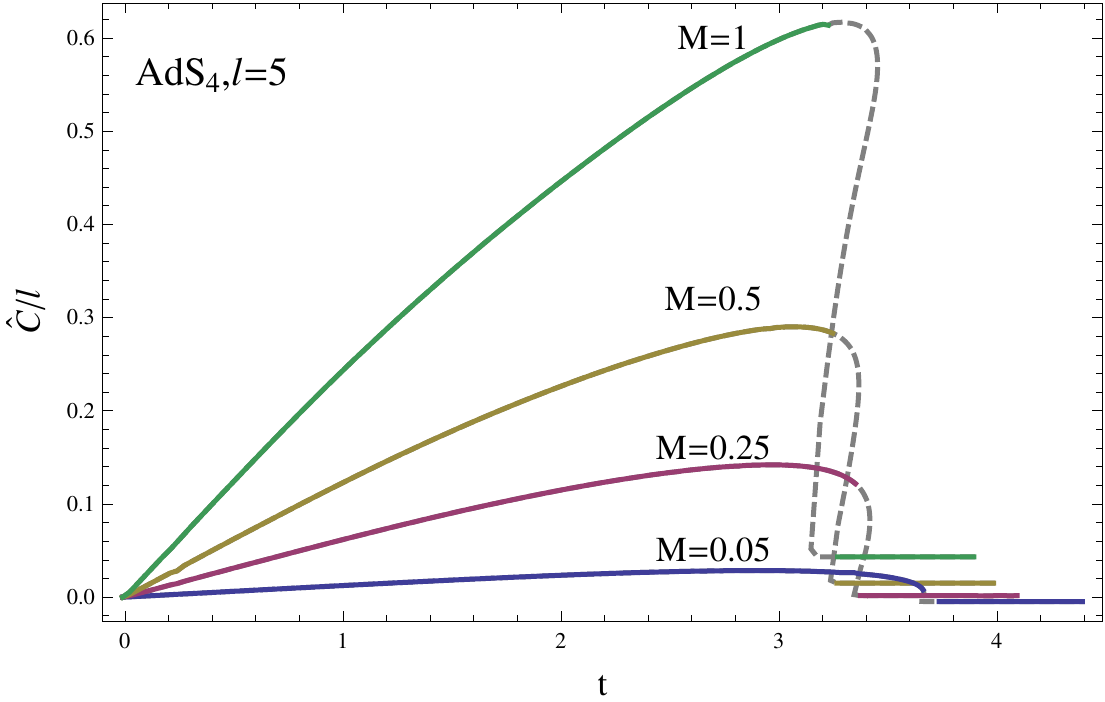}\tabularnewline
\end{tabular}
\par\end{centering}
\caption{\label{fig:ComplexityM}The dependence of the entanglement entropy and
the subregion complexity density $\hat{C}/l$ on $M$. We fix $v_{0}=0.01$
here. Note that there is still a sudden drop of complexity in the evolution when $M=0.05$ in the right lower panel. }
\end{figure}

Now we study the effect of the mass parameter $M$ on the evolution.
We fix the shell thickness $v_{0}=0.01$ here. The numerical results
are shown in Fig. \ref{fig:ComplexityM}. The system evolves from a pure
AdS background to the SAdS black holes with different mass $M$. The maximum
complexity $\hat{C}_{max}$ the system can reach in the evolution
 depends on $M$. For $AdS_{3},l=2$, we get $\hat{C}_{max}/l\propto0.12M$.
For $AdS_{4},l=5$, we get $\hat{C}_{max}/l\propto0.62M$. As we discussed above, $\hat{C}_{max}/l$ is also proportional to $l$. Thus we have
\begin{equation}
\hat{C}_{max}/l \approx f(d) Ml
\end{equation}
where the coefficient $f(d)$ is a function of spacetime dimension.
 Unlike the parameter
$v_{0}$, the increases of $M$ can not change the qualitative behavior
of the evolution, as shown in the lower right  panel. Moreover, the
larger the $M$ is, the sooner the subregion complexity reaches the constant value, as shown
more obviously in the right lower panel.

If we zoom in the final stage of the evolution shown in the
left lower panel in Fig. \ref{fig:ComplexityEntropyT} and Fig. \ref{fig:ComplexityM},
we find that the difference of the complexity between the initial
state and the final state $\hat{C}_{f}$ decreases with $M$ and $l$.
$\hat{C}_{f}$ is more involved in the  right lower panels in Fig. \ref{fig:ComplexityEntropyT}
and Fig. \ref{fig:ComplexityM}.  In this subsection,
we study the dependence of the final  subregion complexity on $M$ and
$l$ in detail.
\begin{figure}
\begin{centering}
\begin{tabular}{cc}
\includegraphics[scale=0.7]{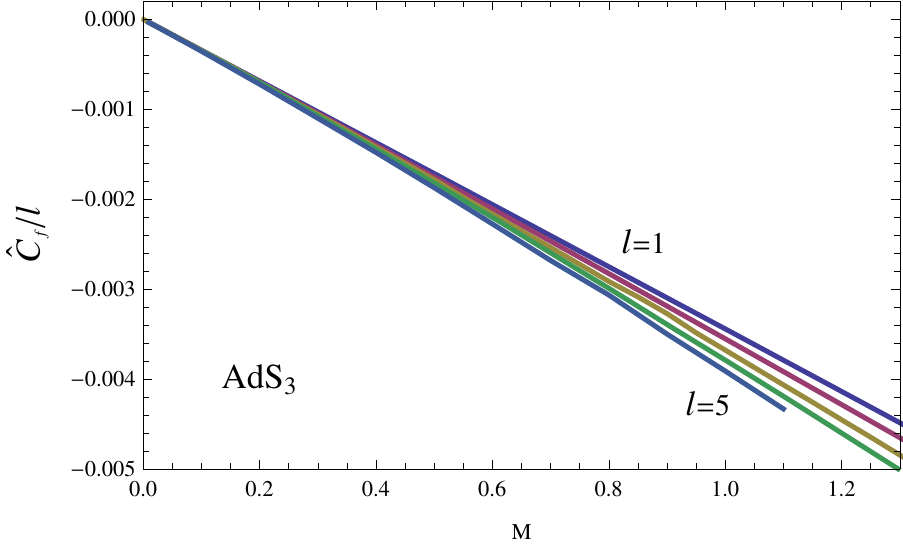} & \includegraphics[scale=0.7]{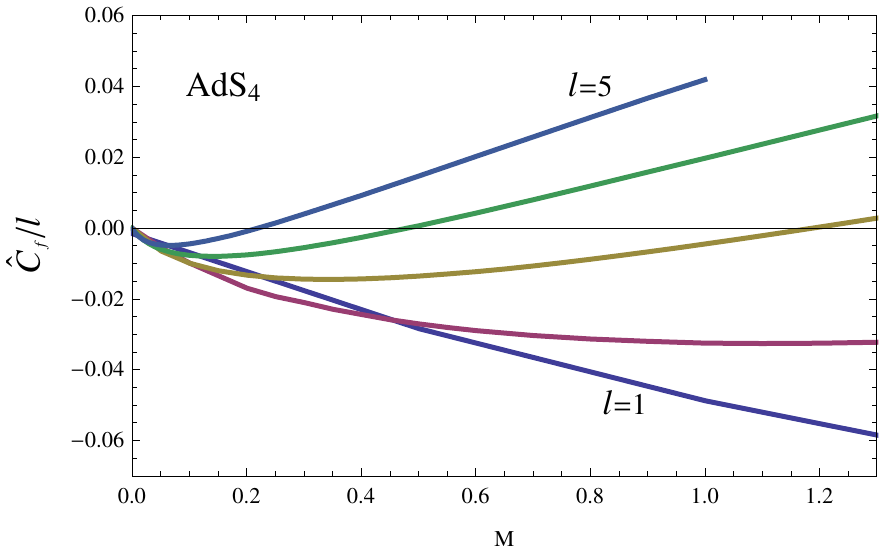}\tabularnewline
\includegraphics[scale=0.7]{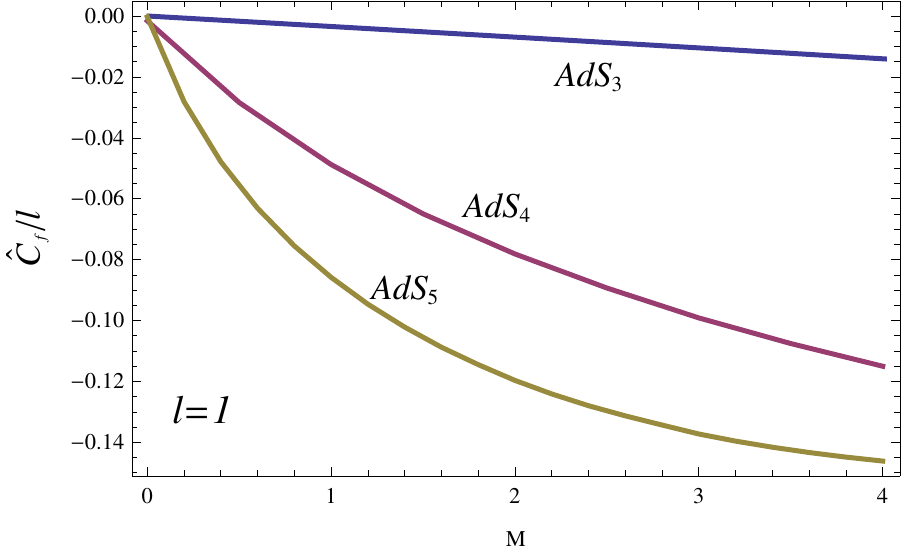} & \includegraphics[scale=0.7]{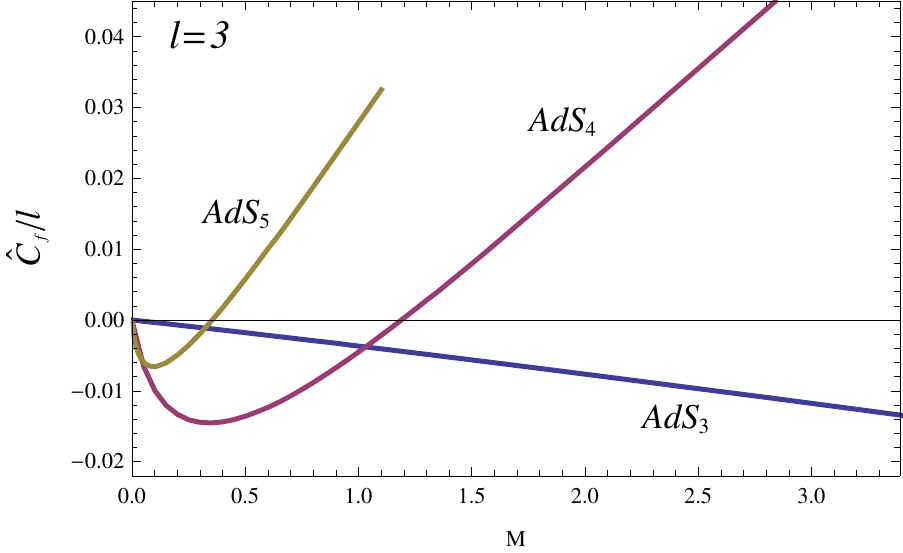}\tabularnewline
\end{tabular}
\par\end{centering}
\caption{\label{fig:ComplexityLengthD}The dependence of subregion complexity
density $\hat{C}_{f}/l$ on the mass parameter $M$.  The upper left panel is for $AdS_{3}$ and $l=1,2,3,4,5$.
The upper right panel is for $AdS_{4}$ and $l=1,2,3,4,5$.}
\end{figure}
For $AdS_{3}$ in the left upper panel of Fig. \ref{fig:ComplexityLengthD}, we see that the complexity
of the final state is always smaller than the initial state. The complexity
density decreases with $M$ linearly for different $l$ and has almost
the same rate $-0.004M$. The situation is more complicated for $AdS_{4}$
shown in the right upper panel of Fig. \ref{fig:ComplexityLengthD}.
The complexity density decreases with almost the same rate for different
$l$ at the beginning. Then it begins to increases with $M$. These
coincide with the behaviors we have found in Fig. \ref{fig:ComplexityEntropyT}
and Fig. \ref{fig:ComplexityM}.

We also compare the dependence of complexity density on the spacetime
dimension. From the left lower panel, we see that the complexity always
decreases with $M$ when the strip size $l$ is not big enough. However,
when the strip size is large, the complexity density would decreases
with $M$ first and then increases almost linearly when $M$ is large
enough in $AdS_{d+1}$ with $d\geq 3$.

\section{Conclusions and discussions}

In this paper, we analyzed the evolution of the subregion complexity  under
a global quench by using numerical method. We considered the situation where
the boundary subregion $\mathcal{A}$ is an infinite strip on a time
slice of the AdS boundary. We followed the subregion CV proposal, which states that the subregion complexity is proportional
to the volume of a codimension-one surface $\Gamma_{\mathcal{A}}$
enclosed by $\mathcal{A}$ and the codimension-two entanglement surface $\gamma_{\mathcal{A}}$
corresponding to $\mathcal{A}$.

We found the following qualitative picture:  the subregion complexity increases at early time after a quench, and after reaching the maximum it  decreases surprisingly to a constant value
at late time.
{ This non-trivial feature is also observed in \cite{Ageev:2018} where the local quench is used to study the subregion CV proposal. The decrease of complexity is also observed in some space like singular bulk gravitational background \cite{Rabinovici:2015,Rabinovici:2018}. It was argued that the decrease of complexity has something to with the entanglement structure. However, as pointed out in \cite{Susskind2014ee}, entanglement is not enough to explain the complexity change. There should be other mechanism for this phenomenon. The evolution of complexity following a quench in free field theory is studied recently \cite{Alves:2018}. It was found that whether the complexity grow or decrease depending on the quench parameters. To compare with the holographic result, the evolution of complexity following a quench in conformal field theory is required.
}


 Another important feature in the subregion complexity under a global quench { we found here} is that when the size of the strip is large enough and the quench is fast enough, in $AdS_{d+1}$ spacetime with $d\geq3$ the evolution of the complexity is discontinuous and there is
 a sudden drop due to the transition of the HRT surface. 

Moreover, at the early time of the evolution,
the growth rates of the subregion complexity densities for the strips of different sizes
 are almost the same. This implies that the complexity growth
is related to the local operators excitations. On the other hand,
for a large enough strip, the subregion complexity grows linearly
with time. If we set the strip size $l\to\infty$, we may expect that
the late time behavior of subregion complexity is linearly increasing. 
However, the large $l \to \infty$ limit should be considered carefully, due to the presence of the holographic
entanglement plateau\cite{Azeyanagi:2007bj,Hubeny:2013gta,Chen:2017ahf}. In this limit, the HRT surface
could be the union of the black hole horizon and the HRT surface for the complementary region.  One has to take into account of this possibility  in discussing the large $l$ limit.
{ Actually, the complexity we considered here for strip with limit  $l \to \infty$ should be reduced to the CV proposal for one-sided black hole. This case has been studied in \cite{Chapman:2018} where it was found that the late time limit of the growth rate of the holographic complexity for the one-sided black hole is precisely the same as that found for an eternal black hole. Thus the complexity for strip with infinite width will not decrease and there will not be a plateau at late time.
}

{
In asymptotic AdS$_3$  black hole case, our results show that the complexity and the corresponding entanglement entropy for subregion will both  keep a  constant approximately if the evolutional time $t\gtrsim l/2$. This can be understood from the thermalization of local states. Ref.~\cite{Cardy:2014rqa} has shown that, for a given quench in 2D CFT, the density matrix of subsystem will be exponentially close to a thermal density matrix if the time is lager than $l/2$. Its correction to thermal state will be suppressed by $e^{-4\pi\Delta_{\min}(t-l/2)/\beta}$. Here $\beta$ is the inverse temperature and $\Delta_{\min}$ is the smallest dimension among those operators which have a non-zero expectation value in the initial state. Thus, we can expect that the complexity and entanglement entropy will suddenly go to their values in corresponding thermal state when the time $t$ is larger than $l/2$. This kind of behavior has been shown clearly in our Fig.~\ref{fig:ComplexityLlong}. The similar behaviours can also be observed in higher dimensional cases, however, the critical time is not $l/2$ but depends on the dimension. This sudden saturation is one characteristic phenomenon in subregion complexity. One can easy see that the critical time of saturation  will approach to infinity if the size of subregion $l$ approaches to infinity.
}

We also analyzed the dependence of the subregion complexity on various
parameters, including the quench speed, the strip size, the black hole
mass and the spacetime dimension. For slow quenches or small strip,
the sudden drop in the subregion complexity evolution disappear such that the complexity evolves continuously. The mass parameter does
no change the qualitative behavior in the evolution when other parameters
are fixed.

Our study can be extended in several directions. Besides the large size limit we mentioned above,  it would be interesting to  consider the evolution
of the subregion complexity under a charge quench or in higher derivative
gravity. It would be certainly interesting to study the subregion complexity by using the CA proposal in order to understand
the holographic complexity better. 

\section*{Acknowledgments}

We thank Davood Momeni for correspondence. B. Chen and W.-M.
Li are supported in part by NSFC Grant No. 11275010, No. 11325522, No.~11335012
and No. 11735001. C.-Y. Zhang is supported by National Postdoctoral
Program for Innovative Talents BX201600005. S.-J. Zhang is supported
in part by National Natural Science Foundation of China (No.11605155).

\end{document}